\documentclass[11 pt]{article}
\usepackage{amssymb}
\usepackage{latexsym}
\usepackage{amsfonts,euscript}
\usepackage{tabularx}
\usepackage{array}
\usepackage{booktabs}
\usepackage{verbatim}
\usepackage{graphicx}
\usepackage{amsmath}
\usepackage{caption}
\usepackage{subcaption}
\usepackage{ulem}
\usepackage{cancel, soul}
\usepackage{url}
 \newtheorem{thm}{Theorem}[subsection]
 
 \newtheorem{lem}[thm]{Lemma}

\usepackage{tikz}
\usetikzlibrary{shapes,arrows}
\usetikzlibrary{matrix,chains,scopes,positioning,arrows,fit}
\usepackage{ctable}
\usepackage{caption}
\tikzstyle{decision} = [diamond, draw, text width=4.5em, text badly centered, node distance=3cm, inner sep=0pt]
\tikzstyle{block} = [rectangle, draw, text width=5em, text centered, rounded corners, minimum height=4em]

\usepackage{booktabs}

\newcommand{\diag}{\operatorname{diag}}
\newcommand{\RR}{\ensuremath{\mathbb{R}}}
\def\R{{\cal R}}
\newcommand{\E}{\operatorname{E}}

\providecommand{\keywords}[1]
{
  \small	
  \textit{Keywords---} #1
}

\newcounter{acounter}

\title{Host movement, transmission hot spots, and vector-borne disease dynamics on spatial networks}
\author{Omar Saucedo$^{1}$\footnote{Corresponding author. E-mail: osaucedo@vt.edu}, Joseph H. Tien$^{2}$\\
\small $^{1}$ Department of Mathematics, Virginia Tech, Blacksburg, VA, USA\\
\small $^{2}$ Department of Mathematics, The Ohio State University, Columbus, OH, USA
}

\begin{document}
\maketitle
\numberwithin{equation}{section}

\begin{abstract}
We examine how spatial heterogeneity combines with mobility network structure to influence vector-borne disease dynamics.  Specifically, we consider a Ross-Macdonald-type disease model on $n$ spatial locations that are coupled by host movement on a strongly connected, weighted, directed graph.  We derive a closed form approximation to the domain reproduction number using a Laurent series expansion, and use this approximation to compute sensitivities of the basic reproduction number to model parameters.  To illustrate how these results can be used to help inform mitigation strategies, as a case study we apply these results to malaria dynamics in Namibia, using published cell phone data and estimates for local disease transmission.  Our analytical results are particularly useful  for understanding drivers of transmission when mobility sinks and transmission hot spots do not coincide.
\end{abstract}


\section{Introduction}
\label{sect:intro}

Vector-borne diseases affect approximately one billion people and account for 17$\%$ of all infectious diseases \cite{Tibayrenc2017}.  Understanding the spatial dynamics of vector-borne diseases is crucial, particularly in the context of increased mobility linking disparate geographic locations \cite{Qiu2013}, changes in mobility patterns due to urbanization, economic development, and globalization, and ecological and environmental changes affecting vector abundance \cite{Sutherst2004, Chen2017}.  The interaction between mobility networks and local transmission characteristics is complex.  Many factors influence local transmission characteristics, including host demography, host density, ecological conditions for the vector (e.g. vegetation, rainfall and temperature, breeding sites), vector abundance, economic resources (e.g. window screens, bed nets), and much more \cite{Mbogo2003spatial,Bousema2010identification}. Spatial locales additionally differ in terms of their connectivity.  Host movement patterns can include the presence of central `hubs' serving to link distinct locales, `source' locales with migration patterns reflecting urbanization, new connections with previously remote locales due to development and changes in land usage, and more \cite{Hay2005urbanization,Matthys2006urban}.  Understanding how these myriad factors combine to influence disease dynamics is challenging, particularly given widespread heterogeneity in habitat suitability and connectivity between locales.
For example, it is possible for movement patterns to allow disease to persist in areas where vector abundance is low \cite{Cosner2015}.   Other work has shown that host movement can lead to disease extinction even in the presence of disease hot spots \cite{Smith2004}.   Thus it is a delicate question how network structure and local disease characteristics combine to affect vector-borne disease dynamics.

There is an extensive literature on using metapopulation models for studying infectious disease dynamics \cite{arino2006,ball2015,citron2021,hanski1999,pei2020,rohani1999,xia2004}, including for vector-borne diseases \cite{Cosner2015, adams2009,barrios2018,cosner2009,gaff2007,gao2014,mpolya2014,mukhtar2020,ruktanonchai2016p,smith2014,torres1997}.  We do not attempt to give a comprehensive review here, but instead simply mention some of the key concepts in the context of the present work.  A fundamental issue is how heterogeneity and connectivity between spatial locations combine to affect disease dynamics, including disease invasion \cite{Cosner2015, bichara2021, gao2012, gao2019, wang2007, zhang2020} and persistence \cite{adams2009, acevedo2015,hasibeder1988}, outbreak size and duration \cite{citron2021,manore2015}, and timing and synchrony between locations \cite{althouse2012,churakov2019}.   Particular considerations for vector-borne diseases include potentially different movement scales between host and vector \cite{smith2014, auger2008}, vector population dynamics including seasonality \cite{aron1982,chitnis2012,hoshen2004} and multi-species interactions \cite{gao2012,reiner2013}, interventions targeting vector and host \cite{agusto2013,demers2020,hughes2013,paton2019}, and challenges for model parameterization \cite{guerra2014,Ruktanonchai2016,tatem2014}.  Many different modeling approaches have been taken for examining these questions, including compartmental models \cite{acevedo2015, bichara2016}, agent-based models \cite{bomblies2014, jindal2017, Mniszewski2014}, network models \cite{kirkland2021, zhao2021, lippi2020}, stochastic models \cite{maliyoni2017,le2018, son2021}, and more.  For a review of approaches for modeling spatial connectivity and mosquito-borne disease dynamics, see \cite{lee2021}.

The primary objective of this manuscript is to present analytical tools for understanding how network structure and local characteristics combine to influence vector-borne disease dynamics.  The impact of heterogeneity and connectivity on the domain reproduction number and control efforts has been studied for specific network types, such as two patch networks \cite{acevedo2015,auger2008}, lattices \cite{demers2020, caraco2001,schwab2018}, star graphs \cite{mpolya2014}, and bipartite graphs \cite{zhang2020, bisanzio2010,iggidr2016}.  One of the contributions of our work is that we derive analytical results for an arbitrary strongly connected graph with arbitrary patch-specific parameters. Specifically, Tien et al \cite{Tien2015} give techniques for computing the domain reproduction number for a network SIR-type model with environmental pathogen movement.  Their methods involve a Laurent series expansion in the next generation matrix approach to computing $\R_0$.  Here we adapt this approach to the vector-borne disease setting.  We consider a network Ross-Macdonald \cite{Smith2012} type model, and derive an approximation for the domain $\R_0$ in terms of an average of the patch $\R_0$ values in isolation, where the average is taken with respect to a probability measure that combines network structure and local pathogen removal rates.  We then use this average to compute sensitivities of the domain $\R_0$ to parameters such as the local transmission and recovery rates.  As an illustration of how these results can be used in practice, we consider malaria in Namibia as a case study.  We use cell phone data from \cite{Ruktanonchai2016} to estimate host movement between health districts, and use published estimates based upon the Malaria Atlas Project for parameterizing local disease characteristics \cite{Gething2011}.  We then compute sensitivities of the domain $\R_0$ for malaria in Namibia based upon these data and our analytical results.  These sensitivities can be useful for guiding intervention strategies.  In particular, our methods give a way to combine information on movement sources and sinks with disease hot spots to assess disease dynamics on the domain.  The Namibia malaria case study also highlights the increasingly available data on mobility patterns and local disease transmission that can be used to apply analytical and computational methods to help understand vector-borne disease dynamics and inform intervention efforts.

The remainder of this article is structured as follows.  In Section \ref{sect:model} we present the modeling framework that will be used throughout.  Section \ref{sect:analysis} gives our analytical results.  Background material on graph theory, random walks on graphs, and key results from \cite{Tien2015} that serve as foundation for our analysis are given in Section \ref{sect:bg}.  After establishing some preliminary results in Section \ref{sect:prelim}, we derive an approximation for the domain $\R_0$ using a Laurent series expansion in Section \ref{GeneralNetworkSection}.  We calculate the sensitivities of this approximation of the domain $\R_0$ to the model parameters in Section \ref{sect:sens}, and apply these results to malaria in Namibia in Section \ref{sect:namibia}.  We conclude with a discussion in Section \ref{sect:discussion}.

\section{Model Description}\label{ModelDescription}
\label{sect:model}

We consider vector-host disease dynamics on $n$ discrete spatial locations.  Disease dynamics within each location follows a Ross-Macdonald-type model \cite{reiner2013,ross1916}, with coupling through host movement.  Host movement follows an Eulerian \cite{Cosner2015,cosner2009,Gueron1996} framework. The resulting system of equations is similar to the Eulerian model considered in \cite{Cosner2015, cosner2009}, but with the inclusion of host vital dynamics, allowing for different host mobility networks depending upon immunological status, and assuming that vectors do not move.  Results by \cite{cosner2009} include establishing global stability of the disease free equilibrium for $\R_0 < 1$ and for the endemic equilibrium when $\R_0 > 1$.

\begin{figure}[!ht]
 \begin{center}

\tikzstyle{block} = [rectangle, draw, fill=blue!20,
    text width=2em, text centered, rounded corners, minimum height=3em, minimum width=3em]

\begin{tikzpicture}[node distance=2.0cm, scale=0.6, every node/.style={scale=1}]

\node[block] (Sn1)                                                      {\Large$S_{i}^{v}$};

\node[] (birth) [below of=Sn1, node distance=1.5cm] {};
\draw[thick, <-] (birth) --node[pos=.5,left]{$\mu_{i}^{v}{S_{i}^{v}}$}(Sn1);

\node[] (death) [left of=Sn1, node distance=1.5cm] {};
\draw[thick, ->] (death) --node[pos=.5,above]{$\Lambda_{i}^{v}$}(Sn1);

\node[] (empty) [right of=Sn1, node distance=4cm] {};
\node[] (empty2) [right of=Sn1, node distance=6cm] {};

\node[block] (Si) [right of=empty, node distance=.15cm]                   {\Large$I_{i}^{v}$};
\draw[thick, ->] (Sn1) --node[pos=.5, below]{$\beta_{i}^{v}\displaystyle\frac{S_{i}^{v}}{N_{i}^h}I_{i}^{h}$}(Si) ;

\node[] (birth4) [below of=Si, node distance=1.5cm] {};
\draw[thick, <-] (birth4) --node[pos=.5, right]{$\mu_{i}^{v}{I_{i}^{v}}$}(Si);

\node[block] (Sn2) [above of=Sn1, node distance=3cm]                        {\Large$S_{i}^{h}$};
\node[] (birth2) [above of=Sn2, node distance=1.5cm] {};
\draw[thick, <-] (birth2) --node[pos=.5,left]{$\mu_{i}^{h}{S_{i}^{h}}$}(Sn2);

\node[](trans)[right of=Sn2, node distance=1.5cm]{};

\node[](trans2)[right of=Sn1, node distance=1.5cm]{};

\draw[thick,dashed, ->] (Si.135)to[out=135, in=315]node[pos=.5,above]{}(trans.45);

\node[] (death) [left of=Sn2, node distance=1.5cm] {};
\draw[thick, ->] (death) --node[pos=.5,above]{$\Lambda_{i}^{h}$}(Sn2);

\node[] (empty3) [right of=Sn2, node distance=4cm] {};

\node[block] (V) [right of=empty3, node distance=.15cm]                 {\Large$I_{i}^{h}$};
\draw[thick, ->] (Sn2) --node[pos=.5, above]{$\beta_{i}^{h}\displaystyle\frac{S_{i}^{h}}{N_{i}^{h}}I_{i}^{v}$}(V) ;

\draw[thick,dashed, ->] (V.235)to[out=235, in=45]node[pos=.5,above]{}(trans2.45);

\node[] (birth3) [above of=V, node distance=1.5cm] {};
\draw[thick, <-] (birth3) --node[pos=.5,left]{$\mu_{i}^{h}{I_{i}^{h}}$}(V);

\draw[thick, ->] (V.205) --node[pos=.75, below]{$\gamma_{i}^{h} {I_{i}^{h}}$}(Sn2.-25) ;

\end{tikzpicture}
\qquad
\tikzstyle{ann} = [draw=none,fill=none,right]
\begin{tikzpicture}[node distance=3.0cm, scale=2, every node/.style={scale=1}]
\tikzstyle{pre}=[<-,shorten <=11pt,>=stealth',thick]
\tikzstyle{post}=[->,shorten >=1pt,>=stealth',semithick]
\tikzstyle{place}=[circle,draw=blue!50,fill=blue!20,thick,
inner sep=0pt,minimum size=6mm]

\node at (0,1) [place]     (waiting) {};
\node[place] (critical) [below of=waiting] {};
\node[place] (leave critical) [right of=critical] {};

edge [post,bend right] (waiting)
\node[place] (enter critical) [left of=critical] {};
\path (enter critical.45) edge [pre,bend right] (waiting.240);
\path (enter critical.90) edge [post,bend left] (waiting.210);

\path (leave critical.90) edge [pre,bend right] (waiting.330);
\path (leave critical.120) edge [post,bend left] (waiting.270);

edge [pre, bend left] (waiting)

\path (critical.-25) edge [pre] (leave critical.205);
\path (critical.10) edge [post] (leave critical.170);
\path (enter critical.-25) edge [pre] (critical.205);
\path (enter critical.10) edge [post] (critical.170);
\end{tikzpicture}

 \end{center}
 \caption{{\it Left:} Model dynamics within each spatial location.  {\it Right:} Schematic of network dynamics for system \eqref{ModelEquations}.  Each node corresponds to a spatial location with SIS-SI dynamics within location.  Nodes are connected by host movement.  Vectors are assumed not to move.}
 \label{flowchart}
\end{figure}
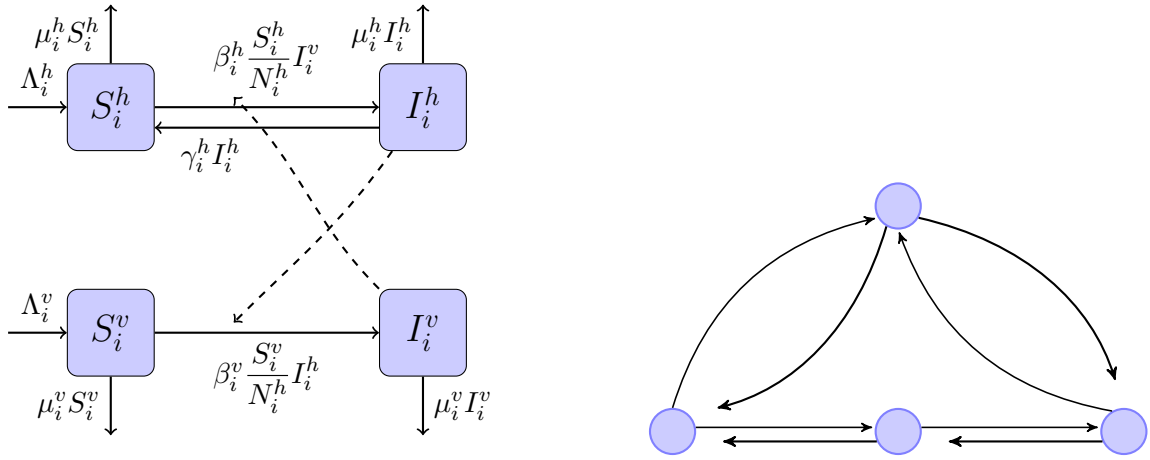

Disease dynamics within a single location follows the flow diagram in Figure \ref{flowchart}.  Let $S_i^{h}$ and $I_i^{h}$ denote the number of susceptible and infectious hosts in location $i$, respectively, and let $S_i^{v}$ and $I_i^{v}$ denote the number of susceptible and infectious vectors.  Let $N_i^{h}$ and $N_i^{v}$ denote the total host and vector populations in location $i$, respectively.  For simplicity, we do not explicitly consider a latent stage for either host or vector.  We also assume there is no infection-derived immunity for hosts.  This is consistent with diseases such as malaria \cite{Doolan2009}, and the SIS-SI system that we use as a building block here has been used previously in malaria modeling studies \cite{Smith2004}.

Let $a_i$ denote the per capita biting rate of vectors on hosts in location $i$, let $b$ correspond to the probability that a bite by an infected vector on a susceptible host results in infection of the host, and let $c$ denote the probability that a bite by a susceptible vector on an infected host results in infection of the vector. We assume an incubation period for vectors of length $\tau$.  Per capita mortality rates for host and vector in location $i$ are denoted by $\mu_i^{h}$ and $\mu_i^{v}$, respectively.  As in the Ross-Macdonald framework, we assume frequency-dependent transmission with transmission parameter from vector to host in location $i$ as $\beta_{i}^{h}=a_i be^{-\mu_{i}^{v}\tau}$, and transmission parameter from host to vector in location $i$ as $\beta_i^{v}=a_i c$.  Infected hosts recover at rate $\gamma_i^{h}$.

Spatial locations are connected through host movement.  We assume that vectors do not move.  This situation may be reasonable on large spatial scales, where connectivity for hosts occurs through automobile and air traffic, but vector dispersal on this spatial scale is rare \cite{adams2009, Stoddard2009, Sumner2017}.  We use an Eulerian framework for movement, corresponding to migration between nodes.  This is in contrast with Lagrangian frameworks for movement, corresponding to commuting between nodes from a distinguished home location.  Both approaches are widely used.  For discussion and comparison of these frameworks, see \cite{cosner2009,hasibeder1988,Gueron1996,Rodriguez2001,esteban2020}.  Eulerian models are often more mathematically tractable.  We will build off the techniques of \cite{Tien2015} to derive analytical results for the vector-host network model considered here.

Let $m^{x}_{ij}$ denote the per capita movement rate of individuals of type $x$ from location $j$ to location $i$, where $x \in \lbrace S, I \rbrace$.  $M^x = \lbrack m^x_{ij} \rbrack$ is thus the weighted adjacency matrix for the mobility network for individuals of type $x$.   Our subsequent analysis will focus on $M^{I}$, the mobility network for infectious hosts.  We will assume throughout that $M^{I}$ is irreducible (Assumption (A\ref{a:irreduc}) below).  This corresponds to a strongly connected mobility network for infectious hosts: for any ordered pair $(j,i)$, there is a path starting from $j$ and terminating at $i$.

We thus have a network of nodes, each with SIS-SI vector-host dynamics and linked to one another through host movement.  Figure \ref{flowchart} shows a schematic.  Note that we are placing minimal constraints on the host mobility network, with the only constraint that the mobility network for $I$ be strongly connected.  Note also that the parameters are allowed to vary between nodes, allowing for heterogeneity in local disease transmission, vector abundance, demography, health care access, and environmental conditions.  Thus the framework we consider is flexible, allowing examination of how network structure and local transmission dynamics interact to affect vector-host disease dynamics on the network. \\

The resulting system of ordinary differential equations is given in \eqref{ModelEquations}:

\begin{equation}
\begin{array}[t]{ll@{}ll}
\dfrac{dS^{h }_{i}}{dt}= \Lambda_{i}^{h}-\beta_{i}^{h} \displaystyle\frac{S_{i}^{h}}{N_{i}^{h}} I_{i}^{v}-\mu_{i}^{h} S_{i}^{h}+\sum_{j=1}^{n} m^{S}_{ij}S_{j}^{h}-\sum_{j=1}^{n}m^{S}_{ji}S_{i}^{h}  + \gamma_i^{h} I_i^{h} \vspace{.2cm} \\
\dfrac{dI^{h }_{i}}{dt}=\beta_{i}^{h} \displaystyle\frac{S_{i}^{h}}{N_{i}^{h}} I_{i}^{v} -(\mu_{i}^{h}+\gamma_{i}^{h}) I_{i}^{h}+\sum_{j=1}^{n}m^{I}_{ij}I_{j}^{h}-\sum_{j=1}^{n}m^{I}_{ji}I_{i}^{h} \vspace{.2cm} \\
\dfrac{dS_{i}^{v}}{dt}= \Lambda_{i}^{v} -\beta_{i}^{v}\displaystyle\frac{S_{i}^{v}}{N_{i}^{h}}I_{i}^{h} -\mu_{i}^{v}S_{i}^{v} \vspace{.2cm} \\
\dfrac{dI_{i}^{v}}{dt}= \beta_{i}^{v}\displaystyle\frac{S_{i}^{v}}{N_{i}^{h}}I_{i}^{h} -\mu_{i}^{v}I_{i}^{v}, \vspace{.5cm}\\
\end{array}
\label{ModelEquations}
\end{equation}
for $i=1,\ldots,n$.

Model parameters and variables are summarized in Table \ref{ParameterValuesModel}.  We make the following assumptions throughout the manuscript:
\begin{list}{{\bf A\arabic{acounter}:~}}{\usecounter{acounter}}

                          \item The parameters $\Lambda_{i}^{h}$, $\Lambda_{i}^{v}$, $\mu_{i}^{h}$, and $\mu_{i}^{v}$  are positive.
                          	\label{a:pos}
                          \item The parameters $\beta_{i}^{h}, \beta_{i}^{v},$ and $\gamma_{i}^{h}$ are non-negative.
                          	\label{a:nneg}
			\item $M^I$ is irreducible.
				\label{a:irreduc}
\end{list}

\begin{table}[h!]
\centering{
\setlength{\tabcolsep}{2pt}
\begin{tabular}{cl}\toprule
Parameter & Description\\
\toprule
$S_{i}^{h}$ & Number of susceptible host in patch $i$.     \\
$I_{i}^{h}$ & Number of infected host in patch $i$.        \\
$R_{i}^{h}$&  Number of recovered host in patch $i$.        \\
$S_{i}^{v}$ & Number of susceptible vector in patch $i$.  \\
$I_{i}^{v}$ & Number of infected vector in patch $i$.   \\
$\Lambda_{i}^{h}$& Host recruitment rate in patch $i$.    \\
$\beta_{i}^{h}$ &  Transmission rate from vector to host in patch $i$  \\
$\mu_{i}^{h}$ & Host death rate in patch $i$.  \\
$\gamma_{i}^{h}$ & Host per capita recovery rate in patch $i$.  \\
$\Lambda_{i}^{v}$ & Vector recruitment rate in patch $i$.  \\
$\beta_{i}^{v}$ & Transmission rate from host to vector in patch $i$.  \\
$\mu_{i}^{v}$ & Vector per capita death rate in patch $i$.   \\
$m^{k}_{ij}$ &  The per capita movement  rate of the host moving from patch $j$ to patch $i$ \\
& from compartment $k=S$, $I$, or $R$.\\
$a$ &  The number of bites on humans, per mosquito, per day.  \\
$b$ &  The transmission efficiency from mosquitoes to humans. \\
    & (i.e. probability of infection from mosquitoes to humans). \\
$c$ & The transmission efficiency from humans to mosquitoes.  \\
    &  (i.e. probability of infection from humans to mosquitoes). \\
$\tau$ &  Vector incubation period.\\
$G_{i}$ & Transfer matrix with respect to the compartments $i=S$, $I$, or $R$. \\
$A$ & Adjacency matrix. \\
$W$ & Diagonal out-degree matrix of A. \\
$L$ & Unnormalized graph Laplacian. \\
$u$ & Strictly positive basis vector of ker $L$ with $\sum_{i=1}^{n} u_{i}=1$. \\
$D$ & Diagonal absorption matrix. \\

\toprule
\end{tabular}
}
\caption{Model parameters for system \eqref{ModelEquations}. The subscript $i$ denotes the $i^{\text{th}}$ patch. } \label{ParameterValuesModel}
\end{table}

\section{Analysis}
\label{sect:analysis}

This section focuses on deriving expressions for $\R_0$ and looking at the sensitivity of $\R_0$ to different parameters.  The technical approach is that introduced in Tien et al (2015), where a Laurent series expansion is given for the fundamental matrix $V^{-1}$ in the next generation matrix.  Central to this is the interplay between movement and removal (``absoption") on the network.  A generalized inverse of the graph Laplacian called the {\it absorption inverse} plays a key role in this analysis \cite{Karly2016}.  Background material on our technical approach is given in Section \ref{sect:bg}.  After some preliminaries (Section \ref{sect:prelim}), we then use this approach in Section \ref{GeneralNetworkSection} to derive a closed form expression that approximates the domain $\R_0$.  This closed form expression can be used to compute approximate sensitivities of $\R_0$ to model parameters; these are given in Section \ref{sect:sens}.
%
\subsection{Absorption, movement, and random walks}
\label{sect:bg}
In this section, for the convenience of the reader, we include some background material on graph theory, random walks on graphs, and the absorption inverse.  Our presentation draws heavily from ideas developed in \cite{Tien2015,Karly2016}.

Let $A \in \RR^{n\times n}$ be the adjacency matrix for a weighted, directed graph $G$ and let $A_{ij}$ define the weight of the arc from $j$ to $i$.  The (unnormalized) graph Laplacian of $G$ is
\begin{equation}
\label{eqn:L}
L = W - A,
\end{equation}
where $W$ is diagonal with $W_{ii} = \sum_{k=1}^n A_{ki}$.  The graph Laplacian encodes a great deal of information about $G$, including the number of components \cite{newman2018networks,Von2007}, enumeration of spanning trees \cite{moon1970}, commute times between nodes via random walk \cite{boley2011commute}, and more.  We specifically mention here two facts that we will use repeatedly: for a strongly connected graph, $\ker L$ is one-dimensional.  Furthermore, a strictly positive basis $u \in \RR^n$ can be found for $\ker L$, with $\sum_{i=1}^n u_i = 1$.  This basis vector corresponds to the stationary distribution of the continuous-time random walk on the graph generated by $L$.  In fact, the entries of $u$ are connected to the spanning trees of $G$ through the matrix tree theorem.  See \cite{Tien2015,moon1970} for additional details.  In addition to the unnormalized graph Laplacian, several other Laplacians exist, and are widely used in studying graph structure.  For an overview, see \cite{Von2007}.

Consider now system \eqref{ModelEquations} and the domain $\R_0$ according to the next generation matrix approach \cite{vandendriessche2002}.  The next generation matrix $FV^{-1}$ involves a matrix $V$ describing the transfer of existing infections.  In \eqref{ModelEquations}, the transfer of existing infection in hosts occurs through either movement or removal (i.e. death or recovery).  The transfer matrix $V$ thus has the following block:
\begin{equation}
\label{eqn:Vsub}
V_{sub} = L + D,
\end{equation}
where $L$ is the graph Laplacian for infected host movement, and $D = \diag\{\mu_{i}^{h} + \gamma_{i}^{h}\}$ corresponds to host removal.  Note that (A\ref{a:pos}) and (A\ref{a:nneg}) imply that $D$ is full rank, and (A\ref{a:irreduc}) together with the preceding discussion give that $L$ has one-dimensional nullspace spanned by $u$.   We will refer to $D$ as the host {\it absorption} matrix.  This appearance of the graph Laplacian plus absorption in the transfer matrix is pervasive in Eulerian modeling frameworks for movement between discrete spatial locations.  See \cite{Tien2015} for an example involving environmental pathogen movement.

For the next generation matrix, we need the inverse of the transfer matrix; hence we will need to invert \eqref{eqn:Vsub}.  As discussed in \cite{Tien2015}, a key point is the relative magnitudes of movement ($||L||$) to absorption ($||D||$).  Let
\begin{equation}
\label{eqn:z}
D = z \bar{D},
\end{equation}
where
\begin{equation}
\label{eqn:Dbar}
\bar{D} = D/\max_{i} D_{ii}.
\end{equation}
Thus $z>0$ is a parameter describing the scale of absorption.  We can consider the effect of varying $z$ while the relative absorption rates between nodes is fixed.

For sufficiently small $z$, $(L + z\bar{D})^{-1}$ can be expressed as a Laurent series.  Particularly important here is the {\it absorption inverse} $L^d$, described in \cite{Karly2016}.  As given in \cite{Karly2016}, for $|z| < 1/\rho(L^d \bar{D})$, we have
\begin{equation}
\label{eqn:laurent}
(L+z\bar{D})^{-1} = \frac{1}{z}U + L^d + \sum_{k=1}^\infty (-z L^d \bar{D})^k L^d,
\end{equation}

\noindent where $U=\dfrac{u\textbf{1}^{T}}{\bar{d}}$. Note that the higher order terms of \eqref{eqn:laurent} depend upon the absorption inverse $L^d$.  The absorption inverse $L^d$ is a generalized inverse of the graph Laplacian, that is closely related to the {\it absorption-scaled graph} $AD^{-1}$.  The absorption inverse thus integrates features of both the network structure ($A$) as well as the absorption rates at the nodes ($D$).  For characterization and description of properties of $L^d$, see \cite{Karly2016}; numerical methods for computing $L^d$ are presented in \cite{benzi2019}.

The Laurent series \eqref{eqn:laurent} can be used to derive approximations to $\R_0$ by truncating the series after a desired number of terms.  For example, for the network model with environmental pathogen movement studied in \cite{Tien2015}, approximating $\R_0$ using the lowest order term in the Laurent series gave
\begin{equation}
\R_0 \approx \E \lbrack \R_{0,i} \rbrack,
\end{equation}
where  $\mathcal{R}_{0,i}$ is the reproduction number of location $i$ in isolation and the expectation is taken with respect to the probability measure $d_i u_i / \bar{d}$, with
\begin{equation}
\label{eqn:dbar}
\bar{d} = \sum_{i=1}^n u_i d_i.
\end{equation}

We shall see that this same probability measure features prominently for the vector-borne disease model studied here.  We thus mention one way to think about this probability measure: $d_i u_i / \bar{d}$ corresponds both to components of a basis for $\ker L^d$, and to the stationary distribution of the continuous-time random walk generated by the Laplacian of the absorption-scaled graph.

\subsection{Preliminaries}
\label{sect:prelim}
As a preliminary to computing $\R_0$ for system (\ref{ModelEquations}), let us first consider the reproduction number for a single node in isolation.  We have the standard result from applying the next generation matrix:

\begin{eqnarray}
\mathcal{R}_{0,i}=\sqrt{\displaystyle\frac{\beta_{i}^{v}\beta_{i}^{h}}{(\mu_{i}^{h}+\gamma_{i}^{h})\mu_{i}^{v}}. }  \label{R0Single}
\end{eqnarray}

Let $L_x$ be the graph Laplacian of the mobility network for individuals of type $x \in \lbrace S, I \rbrace$.  Define
\begin{equation}
\begin{array}{rcl}
G_S &=& L_S + D_\mu, \\
G_I &=& L_I + D_{\mu+\gamma},
\end{array}
\end{equation}
where $D_\mu = \diag \lbrace \mu_i^{h} \rbrace$ and $D_{\mu+\gamma} = \diag \lbrace \mu_i^{h} + \gamma_i^{h} \rbrace$.  In the remainder it will be useful to write $G_I$ in terms of the scaled matrix $\bar{D}_{\mu+\gamma}$ with scaling factor $z$, as described in \eqref{eqn:z}-\eqref{eqn:Dbar}:
\begin{equation}
G_I = L_I + z \bar{D}_{\mu+\gamma}.
\end{equation}

Solving for the disease-free equilibrium (DFE) of the system \eqref{ModelEquations}, we have:

\begin{eqnarray}
\Lambda_{i}^{h}-\mu_{i}^{h} S_{i}^{h}+\displaystyle\sum_{j=1}^{n} m^{S}_{ij}S_{j}^{h}-\displaystyle\sum_{j=1}^{n}m^{S}_{ji}S_{i}^{h}&=&0, \label{Sh} \\
\Lambda_{i}^{v}-\mu_{i}^{v} S_{i}^{v}&=&0. \label{Sv}
\end{eqnarray}

Equation \eqref{Sh} yields $G_{S}S=\Lambda$,  where $S=(S_{1}^{h},\dots, S_{n}^{h})^{T}$ and $\Lambda=(\Lambda_{1}^{h},\dots, \Lambda_{n}^{h})^{T}$.  Notice that (A\ref{a:pos}) implies that $G_{S}$ has the $Z$-sign pattern (i.e. all off-diagonal entries are less than or equal to zero \cite{Horn1985}) with positive diagonal entries, and thus $G_S$ is a nonsingular $M$-matrix (e.g. properties $D_{16}$ and $I_{29}$ in \cite{Berman1994}).
Thus,
\begin{eqnarray}
S^{*}=G_{S}^{-1}\Lambda, \label{Sh2}
\end{eqnarray}
\noindent where $S^{*}=(S_{1}^{h*},\dots, S_{n}^{h*})^{T}$ and $G^{-1}_{S}\geq 0$ by \cite{Berman1994}. \\

From \eqref{Sv} and (A\ref{a:pos}), we have $S^{v*}_{i} = \Lambda_{i}^{v} / \mu_{i}^{v}$.  Therefore, the DFE for system \eqref{ModelEquations} is
\[
(S_{1}^{h*}, 0,  S_{1}^{v*}, 0, S_{2}^{h*}, 0, S_{2}^{v*}, \dots, S_{n}^{h*}, 0, S_{n}^{v*}, 0)^T.
\]

\subsection{Approximating the Domain $\mathcal{R}_{0}$ via Laurent Series Expansion}
\label{GeneralNetworkSection}
We now consider the domain $\R_0$ for system \eqref{ModelEquations} using the next generation matrix \cite{vandendriessche2002}.  Ordering the variables as $(I_{1}^{h}, \dots, I_{n}^{h}, I_{1}^{v}, \dots, I_{n}^{v})$, we have

\begin{equation}
\mathcal{F}=
\left(
\begin{matrix}
 \displaystyle\frac{\beta_{1}^{h}S_{1}^{h}I_{1}^{v}}{N_{1}^{h}}    \cr
 \vdots \cr
 \displaystyle\frac{\beta_{n}^{h}S_{n}^{h}I_{n}^{v}}{N_{n}^{h}}    \cr
 \displaystyle\frac{\beta_{1}^{v}S_{1}^{v}I_{1}^{h}}{N_{1}^{h}}   \cr
 \vdots \cr
 \displaystyle\frac{\beta_{n}^{v}S_{n}^{v}I_{n}^{h}}{N_{n}^{h}}   \cr
\end{matrix}
\right), \quad
\mathcal{V}=
\left(
\begin{matrix}
 (\mu_{1}^{h}+\gamma_{1}^{h})I_{1}^{h}-\displaystyle\sum_{j=1}^{n} m_{1j}I_{j}^{h}+\displaystyle\sum_{j=1}^{n} m_{j1}I_{1}^{h}   \cr
 \vdots \cr
 (\mu_{n}^{h}+\gamma_{n}^{h})I_{n}^{h}-\displaystyle\sum_{j=1}^{n} m_{nj}I_{j}^{h}+\displaystyle\sum_{j=1}^{n} m_{jn}I_{n}^{h}     \cr
 \mu_{1}^{v}I_{1}^{v}   \cr
 \vdots \cr
 \mu_{n}^{v}I_{n}^{v}    \cr
\end{matrix}
\right).
\end{equation}

\noindent Linearizing $\mathcal{F}$ and $\mathcal{V}$ at the DFE gives:
$$ F=
\left(
\begin{matrix}
  0 & D_{\beta}   \cr
  D_{\beta_{v}} & 0   \cr
\end{matrix}
\right),
$$
\begin{center}
$ V=
\left(
\begin{matrix}
  G_{I} & 0  \cr
  0 & D_{\mu_{v}}   \cr
\end{matrix}
\right),
$
\end{center}
where
\begin{eqnarray}
D_{\beta}&=& \diag \lbrace \beta_{1}^{h}, \dots, \beta_{n}^{h} \rbrace,     \vspace{.2cm} \label{F} \\
D_{\beta_{v}}&=& \diag \lbrace \beta_{1}^{v}, \dots, \beta_{n}^{v} \rbrace,   \vspace{.2cm} \label{Fv} \\
D_{\mu+\gamma}&=& \diag \lbrace \mu_{1}^{h}+\gamma_{1}^{h},\dots, \mu_{n}^{h}+\gamma_{n}^{h}  \rbrace,    \vspace{.2cm}\label{DfN} \\
D_{\mu_{v}}&=& \diag \lbrace \mu_{1}^{v},\dots , \mu_{n}^{v} \rbrace.
\end{eqnarray}

\noindent Note that $G_I$ is a nonsingular M-matrix with $G_I^{-1}>0$.  The next generation matrix is thus
\begin{equation}
\label{eqn:FVinv}
 FV^{-1}=
\left(
\begin{matrix}
  0 &   D_{\beta}(D_{\mu_{v}})^{-1} \cr
  D_{\beta_{v}}G_{I}^{-1} & 0   \cr
\end{matrix}
\right).
\end{equation}

We now study the spectral radius of \eqref{eqn:FVinv} using a Laurent series expansion for $G_I^{-1}$.  Recall that $z = \max \lbrace \mu_i^{h} + \gamma_i^{h} \rbrace$ is a measure of how quickly pathogen is removed from the nodes of the network.  Let $L_I^d$ denote the absorption inverse of $L_I$ with respect to the removal (``absorption") rates $\mu_i^{h} + \gamma_i^{h}$.

From \cite{Karly2016}, for $|z| < 1/\rho(L^{d}_{I} \bar{D}_{\mu+\gamma})$ we can express the matrix $G_{I}^{-1}$ as a Laurent series:
\begin{eqnarray}
G_{I}^{-1}=\dfrac{1}{z}U+L^{d}_{I}+\sum_{k=1}^{\infty}(-zL^{d}_{I}\bar{D}_{\mu+\gamma})^{k}L^{d}_{I}, \label{LSGI}
\end{eqnarray}
\noindent where $U=\dfrac{u\textbf{1}^{T}}{\bar{d}}$, $u=(u_{1},\cdots, u_{n})^{T}$ is
a basis for the nullspace of $L_{I}$ with $u>0$ and $\sum_{i=1}^n u_i = 1$, and $\bar{d}$ given by \eqref{eqn:dbar} with $d_i = \mu_i^{h} + \gamma_i^{h}$. Note that $u$ corresponds to the stationary distribution of the random walk on the host mobility network generated by $L_I$. We will refer to $G_I^{-1}$ as the fundamental matrix of the absorbing random walk generated by $G_I$ \cite{Dobrow2016}.

Now consider the eigenvalues of $FV^{-1}$.  We use the following lemma to exploit the block structure of \eqref{eqn:FVinv}.

\begin{lem}\label{lem2}
Let $A$, $B$, $C$, $D \in M_{n\times n}(\mathbb{R})$.  Suppose $A$ is invertible and $AC=CA$.  Then
$$
\det\left(
\begin{matrix}
  A & B \cr
  C & D   \cr
\end{matrix}
\right)=\det(AD-CB). $$
\end{lem}

\noindent {\bf Proof:}  Under the assumption that $A$ invertible, we have

\begin{equation}
\left(
\begin{matrix}
  A^{-1} & 0 \cr
  -C & A   \cr
\end{matrix}
\right)\left(\begin{matrix}  A & B \cr  C & D   \cr \end{matrix} \right)=\left(\begin{matrix}  I & A^{-1}B \cr  -CA+AC & AD-CB   \cr \end{matrix} \right).
\end{equation} \\

\noindent Taking the determinant of both sides and using that $AC=CA$ gives

\begin{equation}
\begin{array}{rcl}
\det \left(
\begin{matrix}
  A^{-1} & 0 \cr
  -C & A   \cr
\end{matrix}
\right)\det\left(\begin{matrix}  A & B \cr  C & D   \cr \end{matrix} \right) &=& \det\left(\begin{matrix}  I & A^{-1}B \cr  0 & AD-CB   \cr \end{matrix} \right) \\
&=& \det (AD-CB).
\end{array}
\end{equation} \\

Next, as the determinant is transpose invariant, we have

\begin{equation}
\det \left(
\begin{matrix}
  A^{-1} & 0 \cr
  -C & A   \cr
\end{matrix}\right)= \det \left(
\begin{matrix}
  A^{-t} & -C^{t} \cr
  0 & A^{t}   \cr

\end{matrix}\right)=\det(A^{-1}A)=1,
\end{equation}
\noindent so
$$
\det\left(
\begin{matrix}
  A & B \cr
  C & D   \cr
\end{matrix}
\right)=\det(AD-CB) . \hspace{2cm} \blacksquare $$ \\
Using Lemma \ref{lem2}, we have that
\begin{eqnarray}
\det(FV^{-1}-\lambda I)&=&\det\left(
\begin{matrix}
  -D_{\lambda} & D_{\beta}D_{\mu_{v}}^{-1} \cr
  D_{\beta_{v}}G_{I}^{-1} & -D_{\lambda}   \cr
\end{matrix}
\right) \\
&=&D_{\lambda}^{2}-D_{\beta_{v}}G_{I}^{-1}D_{\beta}D_{\mu_{v}}^{-1},
\label{R02}
\end{eqnarray}
so the eigenvalues of the matrix $D_{\beta_{v}}G_{I}^{-1}D_\beta D_{\mu_v}^{-1}$ are the square of the eigenvalues of $FV^{-1}$.
As the matrix $D_{\beta_{v}}G_{I}^{-1}D_\beta D_{\mu_v}^{-1}$ is similar to the matrix $G_{I}^{-1}D_\beta D_{\mu_v}^{-1}D_{\beta_{v}}$, it suffices to find the eigenvalues of the matrix $G_{I}^{-1}D_\beta D_{\mu_v}^{-1}D_{\beta_{v}}$.

We now use the Laurent series \eqref{LSGI} for $G_I^{-1}$ for finding approximate eigenvalues for  $G_{I}^{-1}D_\beta D_{\mu_v}^{-1}D_{\beta_{v}}$.  Let
\begin{equation}
\label{eqn:barD}
\hat{D}=D_\beta D_{\mu_v}^{-1}D_{\beta_{v}}=\diag
\Big \lbrace \displaystyle\frac{\beta_{1}^{h}\beta_{1}^{v}}{\mu_{1}^{v}}, \cdots, \displaystyle\frac{\beta_{n}^{h}\beta_{n}^{v}}{\mu_{n}^{v}} \Big \rbrace,
\end{equation}
and consider the approximation for $G_I^{-1}$ using only the lowest order term in \eqref{LSGI}.  Then we have
\begin{equation}
G_{I}^{-1}\hat{D} \approx \dfrac{1}{z}U\hat{D}.
\end{equation}

Note that $U$ is rank 1 and $\hat{D}$ has full rank,
implying that $\frac{1}{z}U\hat{D}$ has only one nonzero eigenvalue, which we can directly compute:

\begin{equation}
\begin{array}{rcl}
  \dfrac{1}{z} U\hat{D}u &=& \dfrac{1}{z}\dfrac{u\textbf{1}^{T}}{\bar{d}}\hat{D}u \vspace{.1in}\\
  &=& \dfrac{1}{z}\dfrac{1}{\bar{d}}u\textbf{1}^{T}(\hat{D}_{11}u_{1}, \cdots, \hat{D}_{nn}u_{n})^{T} \vspace{.1in}\\
   &=& \dfrac{1}{z}\left(\dfrac{\displaystyle\sum_{i=1}^{n}\hat{D}_{ii}u_{i}}{\bar{d}}\right) u.
\end{array}
\end{equation}

Thus, the spectral radius of $D_{\beta_{v}}G_{I}^{-1}D_\beta D_{\mu_v}^{-1}$ is approximately  $ \displaystyle \frac{1}{z}(\sum_{i=1}^{n}\hat{D}_{ii}u_{i})/\bar{d}$.

Substituting \eqref{eqn:barD} and \eqref{eqn:dbar} for $\hat{D}$ and $\bar{d}$, respectively, and using \eqref{R0Single}, we have the following lowest order approximation to the domain reproduction number:

\begin{equation}
\label{NetworkR0}
\begin{array}{rcl}
\R_0 &\approx& \dfrac{1}{\sqrt{z}}\sqrt{\displaystyle\sum_{i=1}^n  \ \hat{D}_{ii} u_i \frac{1}{\bar{d}}}  \vspace{.1in}\\
&=& \dfrac{1}{\sqrt{z}}\sqrt{\displaystyle\sum_{i=1}^n  \ \frac{\beta_i^{h} \beta_{i}^{v} u_i}{\mu_{i}^{v}} \frac{1}{\bar{d}}} \vspace{.1in} \\
&=& \dfrac{1}{\sqrt{z}}\sqrt{\displaystyle\sum_{i=1}^n  \ \frac{\beta_i^{h} \beta_{i}^{v}}{\mu_{i}^{v}(\mu_i^{h}+\gamma_i^{h}) } d_i u_i \frac{1}{\bar{d}}} \vspace{.1in} \\
&=& \dfrac{1}{\sqrt{z}}\sqrt{\displaystyle\sum_{i=1}^n  \ \lbrack \R_{0,i} \rbrack^2 d_i u_i \frac{1}{\bar{d}}} \vspace{.1in} \\
&=& \dfrac{1}{\sqrt{z}}\sqrt{\E \lbrack \R_{0,i}^2 \rbrack} \vspace{.1in} \\
&:=& \hat{\R}_0,
\end{array}
\end{equation}
where $d_{i}=\mu_i^{h}+\gamma_{i}^{h}$ and the expectation of $\mathcal{R}_{0,i}^2$ is with respect to the probability measure $d_{i}u_{i}/\bar{d}$.  As pointed out in \cite{Karly2016}, this probability measure corresponds to the stationary distribution of the random walk generated by the Laplacian for the `absorption-scaled graph'.  Expression \eqref{NetworkR0} is the analogue to the approximation obtained in \cite{Tien2015} for `SIWR' systems \cite{tien2010} coupled by environmental pathogen movement, and shows how network structure (through the $u_i$ in the probability measure) and local node characteristics (the node-specific removal rates $d_i$, and the node-specific reproduction numbers) combine to shape the domain $\R_0$.

\subsection{Sensitivity analysis}
\label{sect:sens}
The closed form approximation for the domain $\mathcal{R}_0$ in \eqref{NetworkR0} allows us to probe how network structure and local disease characteristics combine to affect disease dynamics.  In this section we use \eqref{NetworkR0} to examine the sensitivity of the domain $\mathcal{R}_0$ to model parameters.  This can help in evaluating the potential impact of intervention strategies.

For sufficiently small $z$, we can approximate the sensitivity of $\mathcal{R}_0$ to parameter $p$ by differentiating the expression in \eqref{NetworkR0} with respect to $p$ and multiplying by the normalizing term $\dfrac{p}{\mathcal{R}_{0}}$ as in \cite{chitnis2008determining} yields the following scaled sensitivities:

\begin{eqnarray}
\dfrac{\partial\mathcal{R}_{0}}{\partial\beta_{i}^{h} }\dfrac{\beta_{i}^{h}}{\mathcal{R}_{0}}  &\approx& \dfrac{\partial \mathcal{\hat{R}}_{0}}{\partial\beta_{i}^{h} }\dfrac{\beta_{i}^{h}}{\mathcal{\hat{R}}_{0}}
   = \dfrac{1}{2z}\frac{1}{\sqrt[3]{\mathcal{\hat{R}}_{0}}}[\mathcal{R}_{0,i}]^2\dfrac{d_{i}u_{i}}{\bar{d}}, \label{EqSenBeta}
\end{eqnarray}

\begin{eqnarray}
\dfrac{\partial\mathcal{R}_{0}}{\partial\beta_{i}^{v} }\dfrac{\beta_{i}^{v}}{\mathcal{R}_{0}}   &\approx&\dfrac{\partial\mathcal{\hat{R}}_{0}}{\partial\beta_{i}^{v} }\dfrac{\beta_{i}^{v}}{\mathcal{\hat{R}}_{0}}  = \dfrac{1}{2z}\frac{1}{\sqrt[3]{\mathcal{\hat{R}}_{0}}}[\mathcal{R}_{0,i}]^2\dfrac{d_{i}u_{i}}{\bar{d}}, \label{EqSensBetav}
\end{eqnarray}

\begin{eqnarray}
 \dfrac{\partial\mathcal{R}_{0}}{\partial\gamma_{i}^{h} }\dfrac{\gamma_{i}^{h}}{\mathcal{R}_{0}}  &\approx&   \dfrac{\partial\mathcal{\hat{R}}_{0}}{\partial\gamma_{i}^{h} }\dfrac{\gamma_{i}^{h}}{\mathcal{\hat{R}}_{0}}
   = -\frac{1}{2}\sqrt{\mathcal{\hat{R}}_{0}} \Big(\dfrac{\gamma_{i}^{h} u_{i}}{\bar{d}}\Big), \label{EqSensGamma}
\end{eqnarray}

\begin{eqnarray}
\dfrac{\partial\mathcal{R}_{0}}{\partial \mu_{i}^{v} }\dfrac{\mu_{i}^{v}}{\mathcal{R}_{0}}  &\approx&   \dfrac{\partial\mathcal{\hat{R}}_{0}}{\partial \mu_{i}^{v} }\dfrac{\mu_{i}^{v}}{\mathcal{\hat{R}}_{0}}
   = -\dfrac{1}{2z} \frac{1}{\sqrt[3]{\mathcal{\hat{R}}_{0}}} [\mathcal{R}_{0,i} ]^{2}\dfrac{d_{i}u_{i}}{\bar{d}}. \label{EqSensFv}
\end{eqnarray}

Each of the expressions in \eqref{EqSenBeta} - \eqref{EqSensFv} shows how network structure and local dynamics combine to shape the $\R_0$ sensitivities.  In particular, we see that the relative sensitivities between nodes of $\R_0$ to a given parameter are determined either by $\R_{0,i}^2 d_i u_i$ (for the sensitivities to the transmission parameters $\beta_i^{h} , \beta_{i}^{v} $, and to the vector mortality rates $\mu_{i}^{v} $) or $\gamma_i^{h}  u_i$ (for the sensitivity to the host recovery rates $\gamma_i^{h}$).

Determining effective intervention strategies in the context of heterogeneous local transmission and complex mobility networks is challenging.  The sensitivities \eqref{EqSenBeta} - \eqref{EqSensFv} can be useful for evaluating intervention efforts such as vector elimination (e.g. increasing $\mu_{i}^{v}$ in targeted locations), bed net distribution (e.g. decreasing $\beta_{i}^{h} $ and $\beta_{i}^{v} $), or chemotherapy following infection (e.g. increasing $\gamma_{i}^{h} $) \cite{Sutherst2004}.  We illustrate this first in Section \ref{sect:toy} for a toy example, to show the interaction between transmission hot spots and mobility sources and sinks.  We then turn to an empirical case study of malaria dynamics in Namibia in Section \ref{sect:namibia}.

\begin{figure*}[htb]
    \centering
\begin{subfigure}{0.25\textwidth}
  \includegraphics[width=\linewidth]{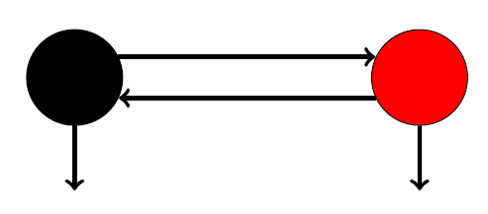}
  \caption{}
  \label{fig:case1}
\end{subfigure}
\begin{subfigure}{0.25\textwidth}\renewcommand{\thesubfigure}{c}
  \includegraphics[width=\linewidth]{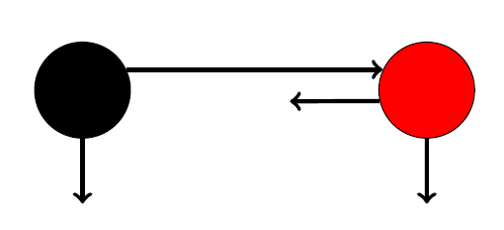}
  \caption{}
  \label{fig:case2}
\end{subfigure}\hfil

\medskip
\begin{subfigure}{0.25\textwidth}\renewcommand{\thesubfigure}{b}
  \includegraphics[width=\linewidth]{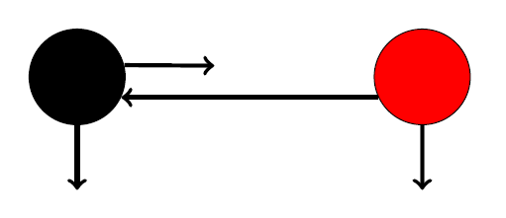}
  \caption{}
  \label{fig:case3}
\end{subfigure}
\begin{subfigure}{0.25\textwidth}
  \includegraphics[width=\linewidth]{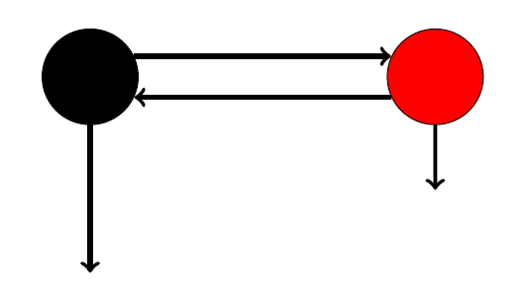}
  \caption{}
  \label{fig:case4}
\end{subfigure}

\medskip
\begin{subfigure}{0.25\textwidth}
  \includegraphics[width=\linewidth]{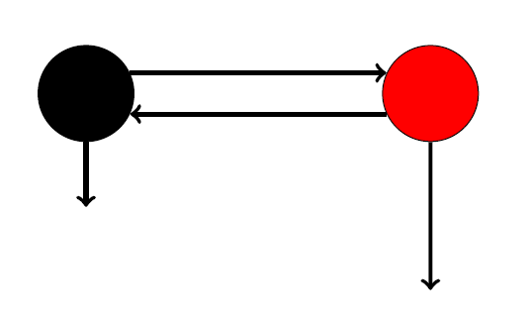}
  \caption{}
  \label{fig:case5}
\end{subfigure}
\caption{Toy example illustrating the relationship between the patch specific reproduction number, network risk, and host absorption rate. Node 1 = black, node 2 = red.  Red signifies a transmission hot spot.  For all cases, $\mathcal{R}_{0, 1} <1< \mathcal{R}_{0, 2}$. We set up five scenarios in which we change the movement rates $m_{ij}$ and the effective network structure $d_{i}u_{i}$. In the cases where the effective network structure terms are not equal, we assume the transmission rates are higher in node 2 in order to preserve the assumption of node 2 being a transmission hot spot.  (a) movement rates are equal $(m_{12}=m_{21})$ and the effective network structure terms are equal $(d_{1}u_{1}=d_{2}u_{2})$, (b) movement rates between the nodes are different $(m_{12}>m_{21})$ and the effective network structure terms are distinct  $(d_{1}u_{1}>d_{2}u_{2}, d_{1}=d_{2})$ , (c) movement rates between the nodes are different $(m_{12}<m_{21})$ and the effective network structure terms are distinct $(d_{1}u_{1}<d_{2}u_{2}, d_{1}=d_{2})$ (d) movement rates are equal $(m_{12}=m_{21})$ and the effect network structure terms are distinct $(d_{1}u_{1}>d_{2}u_{2}, u_{1}=u_{2})$ (e) movement rates are equal $(m_{12}=m_{21})$ and the effect network structure terms are distinct $(d_{1}u_{1}<d_{2}u_{2}, u_{1}=u_{2}).$ }
\label{fig:ToyExample}
\end{figure*}

\subsubsection{Toy example}
\label{sect:toy}


A practical consideration regarding vector-borne disease dynamics is the role that mobility sources, mobility sinks, and transmission hot spots play in driving transmission.  In the `strongly coupled' regime where the Laurent series is well-approximated by the first term, the sensitivities \eqref{EqSenBeta}-\eqref{EqSensFv} can be used to disentangle this, as the sensitivity expressions include terms reflecting network structure ($u_i$), local transmission dynamics ($\R_{0,i}$), and host removal (	``absorption rates", $d_i$).  The $d_i$ appear in the sensitivities \eqref{EqSenBeta}, \eqref{EqSensBetav}, and \eqref{EqSensFv} as the product $d_i u_i$, which can be viewed as an {\it effective network structure} term taking absorption rates into account.  Indeed, the $d_i u_i$ correspond to the stationary distribution of a random walk on the {\it absorption-scaled graph} introduced in \cite{Tien2015}. We illustrate how all these characteristics influence $\mathcal{\hat{R}}_{0}$ (and thus, for sufficiently small $z$, $\R_0$) with a set of toy examples.

Consider a two-node network, letting the local $\R_{0,i}$ values be fixed and taking node two as a transmission hot spot ($\R_{0,2}>1$). A schematic is shown in Figure \ref{fig:ToyExample}, with red denoting the transmission hot spot (node 2), horizontal arrows corresponding to movement between nodes, and vertical arrows corresponding to absorption.  We assume that as the absorption rates vary (for example, between (a) and (d)), the transmission rates compensate in order to keep $\R_{0,i}$ fixed. We now consider several scenarios.  In the first three cases, we vary the movement rates and the effective network structure terms by changing $u_{i}$.  In (a), we have a network with equal movement rates $m_{12} = m_{21}$ and equal absorption rates between the nodes. Thus the effective network structure terms $d_i u_i$ are equal, and according to \eqref{EqSenBeta}, \eqref{EqSensBetav}, and \eqref{EqSensFv}, $\hat{\R}_0$ is more sensitive to the transmission parameters $\beta_{i}^{h}, \beta_{i}^{v}$ and vector mortality rate ($\mu_{i}^{v}$) in  the transmission hot spot (node two).  However, \eqref{EqSensGamma} indicates that the sensitivity of $\hat{\R}_0$ to the host recovery rates $\gamma_i^{h}$ are approximately equal between nodes.   These results suggest that for balanced networks (i.e. networks where all the patches have equal net inflow and net outflow) with equal absorption rates, targeting transmission hot spots for interventions affecting transmission rates and vector mortality will lead to the largest reductions in $\hat{\R}_0$.  In contrast, interventions that increase host recovery rates (for example, through increased treatment rates) will lead to comparable reductions in $\hat{\R}_0$ regardless of where the intervention is applied.

In (b), we consider an unbalanced network with $m_{12} > m_{21}$ and $d_{1}=d_{2}$. In the language of \cite{tatem2014},  node one is a mobility {\it sink}, and node two is a mobility {\it source}.  This situation corresponds to $u_1 > u_2$, and the transmission hot spot corresponds to a mobility source on the absorption-scaled graph. Whether $\hat{\R}_0$ is more sensitive to the transmission ($\beta_i^{h} , \beta_{i}^{v} $) and vector mortality ($\mu_{i}^{v} $) parameters for node one versus node two depends upon the specific numerical values of $u_i$ and $\R_{0,i}$.  Thus further details are necessary for determining where to target interventions affecting $\beta_i^{h} , \beta_{i}^{v} $, and $\mu_{i}^{v} $.

In (c), we switch the mobility rates so that node one is the mobility source, and node two the mobility sink (i.e. $m_{12} < m_{21}$) and again let $d_{1}=d_{2}$. Here the transmission hot spot corresponds to a mobility sink on the absorption-scaled graph. $\hat{\R}_0$ is more sensitive to the parameters in node two compared to node one since we have that $\mathcal{R}_{0,1}u_{1}<\mathcal{R}_{0,2}u_{2}$ for the transmission and removal parameters. Thus interventions at nodes that are simultaneously mobility sinks and transmission hot spots are likely to lead to relatively large reductions in domain $\hat{\R}_0$ when targeting transmission and removal terms.

In (d) and (e) we set $m_{12} = m_{21}$, but let $d_1 \neq d_2$. It is possible for the transmission hot spot to correspond to a mobility sink on the absorption-scaled graph, in which case $\hat{\R}_0$ is more sensitive to all the parameters in a transmission hot spot.  This case is shown in (d).  It is also possible for the transmission hot spot to correspond to a mobility source on the absorption-scaled graph.  In this case the relative sensitivities of $\hat{\R}_0$ to $\beta_i^{h} , \beta_{i}^{v} $, and $\mu_{i}^{v} $ in node one versus two depend upon the specific numerical values for $\R_{0,i}$ and $d_i u_i$.

\section{Case Study: Malaria in Namibia}
\label{sect:namibia}

We turn now to a case study of malaria in Namibia to illustrate how the techniques we have developed can be applied in practice.  Data for this section are described in \cite{Ruktanonchai2016}.  Notably, \cite{Ruktanonchai2016} provides both detailed empirical data on host mobility through cell phone records, as well as parameter estimates for local disease transmission characteristics through the Malaria Atlas Project.  These are precisely the types of data required for applying our methods for approximating the domain $\R_0$ and examining its sensitivities to model parameters for evaluating intervention strategies.

\subsection{Mobility network}
Ruktanonchai et al \cite{Ruktanonchai2016} use anonymized cell phone records involving 9 billion communications sent by 1.19 million unique SIM cards in 2010 to estimate mobility patterns in Namibia.  Locations were determined to the health district level (level 2 shapefile for Namibia given by Database of Global Administrative Areas) based upon cell tower corresponding to last communication for each day.  Movement between health districts was estimated from changes in cell tower locations between communications, and resident health districts for each SIM card were estimated from most frequent location over a one year period.  Communication records were then aggregated across SIM cards to estimate a mixing matrix $P=(p_{ij})_{n\times n}$ giving the proportion of time that a resident of $j$ spent in $i$. The matrix $P$ is a non-negative matrix with column sums equal to 1. For additional details, see \cite{Ruktanonchai2016}.

Note that the cell phone data represent aggregate data with respect to immunological status.  No distinction is made between infected and non-infected individuals in the data.  As the majority of the data likely come from individuals whose movement is not affected by illness, these data are likely representative of movement patterns of healthy individuals.  Illness has the potential to significantly impact movement patterns \cite{lin2016}.  In the absence of movement data specific to immunological status, we nevertheless use the aggregate cell phone data here to estimate the mobility network for infected individuals.  Further understanding of how illness affects movement is an important area for future work.

The mixing matrix estimated by \cite{Ruktanonchai2016} is suitable for a Lagrangian modeling framework \cite{cosner2009}.  To adapt to the Eulerian framework used in \eqref{ModelEquations}, we relate the two modeling frameworks through the expected `residence times' of the two frameworks.  We briefly describe our approach below; further details on this approach and comparison of the Lagrangian and Eulerian frameworks in the context of vector-borne disease is given in \cite{esteban2020}.

We wish to use the mixing matrix $P$ to estimate the adjacency matrix $M^I$ in \eqref{ModelEquations}.  Specifically, we do so by matching the expected times that individuals from $j$ spend in $i$ in both models.  For system \eqref{ModelEquations}, the $i,j$ entry of $(L+D_{\mu+\gamma})^{-1}$ corresponds to the expected time an (infectious) host in $j$ spends in $i$, while the empirically estimated $p_{ij}$ corresponds to the proportion of time a resident of $j$ spends in $i$.  Multiplying this proportion by the infectious period $1/(\mu_{j}^{h}+\gamma_{j}^{h})$ gives the expected time that an infectious resident of $j$ spends in $i$.  Equating these two expressions gives
\begin{equation}
\label{eqn:consistent}
(L+D_{\mu+\gamma})^{-1} = PD_{\mu+\gamma}^{-1}.
\end{equation}

Given $P$ and $D_{\mu+\gamma}$, it remains to estimate $L$.  We do so by using the Frobenius norm to treat this as a constrained least squares problem, in the case where $P$ is invertible (as is indeed the case for the Namibia cell phone data).
Suppose that $P$ invertible, and let $\hat{P}^{-1} = PD_{\mu+\gamma}^{-1} $.  Then with \eqref{eqn:L} we have

\begin{equation}
\label{MinProblem} 
\begin{array}{rcl}
\min_{A\geq 0} ||(L+D_{\mu+\gamma})^{-1}- \hat{P}^{-1}||_{F}  &=& \min_{A\geq 0} || L + D_{\mu+\gamma} - \hat{P}||_{F} \\
&=& \min_{A \geq 0} ||W-A+D_{\mu+\gamma}-\hat{P} ||_F \\
&=& \min_{A \geq 0} ||W-A-(D_{\mu+\gamma}P^{-1}-D_{\mu+\gamma})||_F \\
&=& \min_{A \geq 0} ||W-A-D_{\mu+\gamma}(P^{-1}-I)||_F.
\end{array}
\end{equation}

\noindent In the case where all of the absorption rates are the same, $D_{\mu+\gamma}$ is equal to a scalar multiple of the identity and \eqref{MinProblem} can be written as
\begin{eqnarray}
\min_{\hat{A} \geq 0} ||\hat{W}-\hat{A}-(P^{-1}-I)||_F,
\label{eqn:min}
\end{eqnarray}
\noindent where $W=\hat{W}D_{\mu+\gamma}$ and $A=\hat{A}D_{\mu+\gamma}$.  \\

\begin{figure}[htbp]
\centering
\scalebox{0.75}{\includegraphics{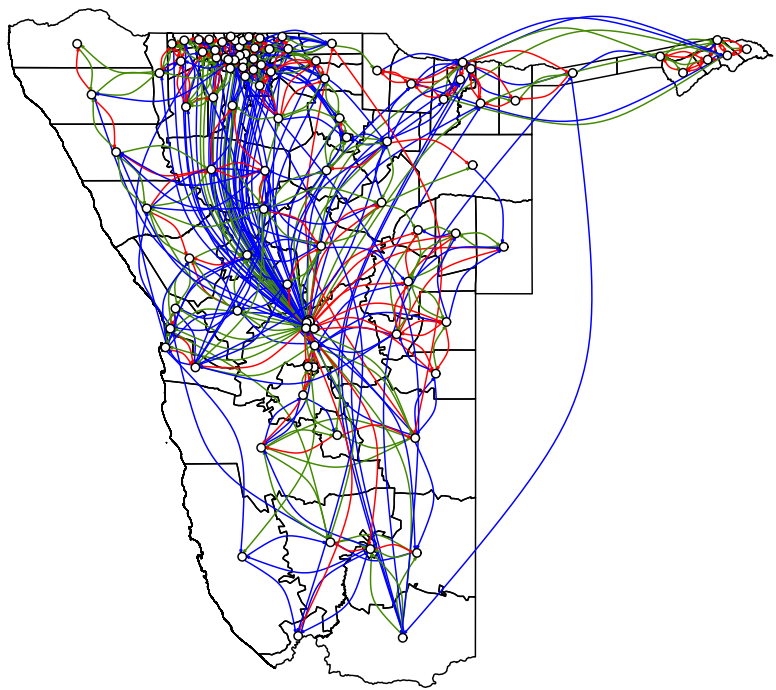}} 
\caption{The estimated connectivity network for Namibia. The distinct color lines on the map represent the level of connectivity between each region.  The blue lines indicate a weak connectivity $(A_{ij}<0.019)$, the green lines represent an intermediate level of connectivity $(0.019<A_{ij}<0.039)$, and the red lines denote strong connectivity $(A_{ij}>0.039)$. To avoid an overcrowded graph, we show values $A_{ij}>0.01$. }
\label{NambiaNetworkMap}
\end{figure}

Numerical solutions to \eqref{eqn:min} were computed in {\sc Matlab R2018a} using the constrained linear least squares function \texttt{lsqlin}.  The resulting estimated adjacency matrix for Namibia based upon the cell phone data is shown in Figure \ref{NambiaNetworkMap}.

\begin{figure}[h]
  \centering
\scalebox{0.5}{\includegraphics{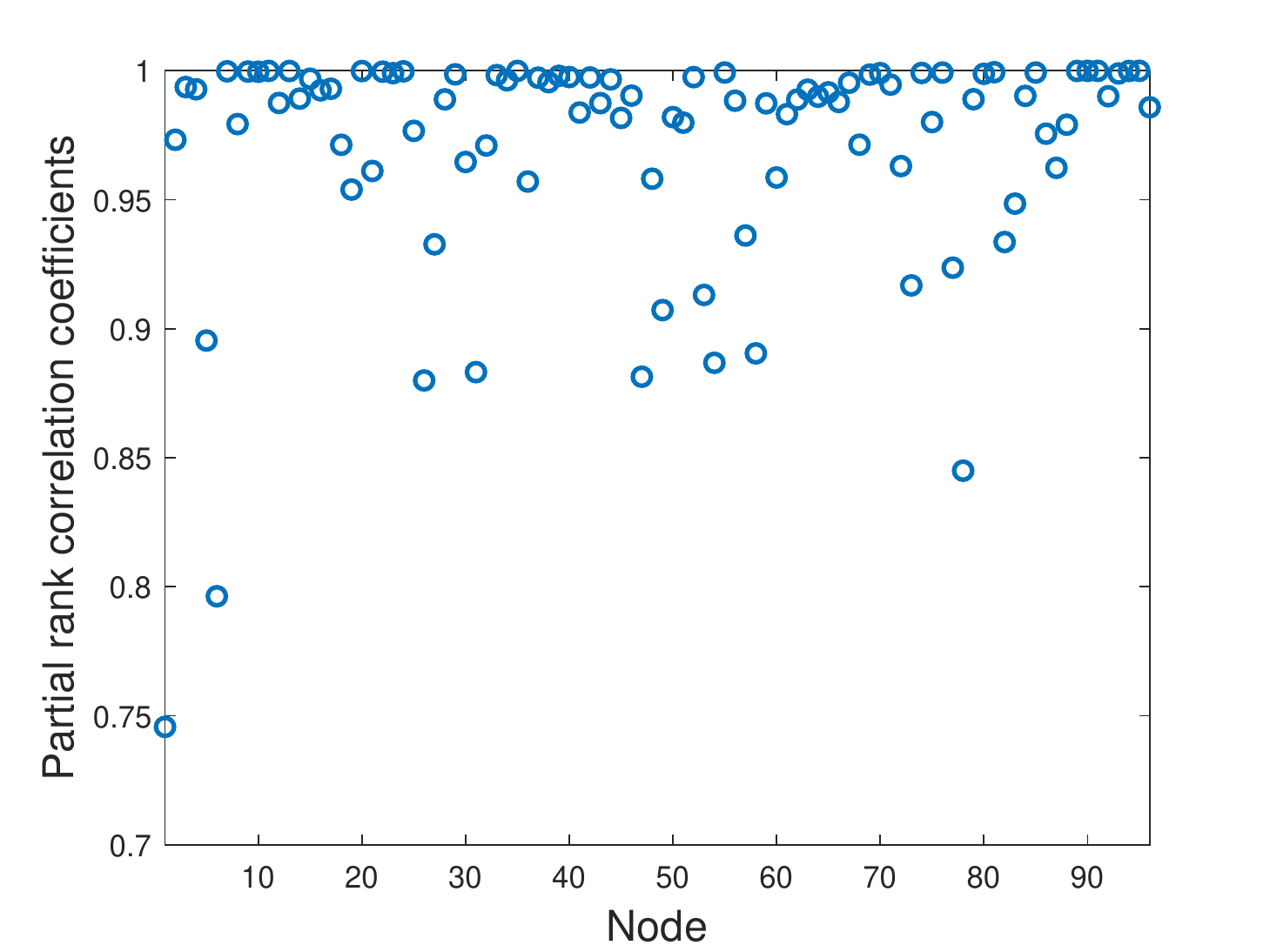}}
  \caption{Partial rank correlation coefficients (PRCC) between the empirically observed mixing matrix based upon cell phone data from \cite{Ruktanonchai2016}, and residence times $(L+D)^{-1}$ for system \eqref{ModelEquations}.  The majority of the PRCC values are above 0.90.}
\label{PartialRCC}
\end{figure}

As discussed in \cite{esteban2020}, \eqref{eqn:min} may not have a solution corresponding to zero.  In the language of \cite{esteban2020}, such a situation is said to be {\it inconsistent}, and this is the case for the Namibia data.  To examine how closely the estimated adjacency matrix $M^I$ matches the original cell phone data, we consider the partial rank correlation coefficients between the entries of $(L+D_{\mu+ \gamma})^{-1}$ and $\hat{P}^{-1}$ \cite{Marino2008}. The resulting partial rank correlation coefficients are shown in Figure \ref{PartialRCC} (computations were done using the \texttt{partialcorr} function in {\sc Matlab}). From Figure \ref{PartialRCC}, we observed that over 90\% of the PRCCs are above the value 0.90.

\subsection{Local transmission parameters}
Parameters for each health district were based upon estimates from \cite{Ruktanonchai2016}, who used data from the Malaria Atlas Project to parameterize a Ross-Macdonald SIS model in a Lagrangian framework for host movement.  Specifically, Ruktanonchai et al give estimates of 0.3 for the human feeding rate of mosquitoes per day, 0.1 for the probability of transmitting from mosquito to human, 0.214 for the probability of transmitting from human to mosquito, 10 days for the incubation period, 0.1 per day for the death rate for mosquito, and $1/150$ per day for the recovery rate of humans \cite{Ruktanonchai2016}.
We additionally estimate the host mortality rate to be $1/23002$ per day from census data for Namibia \cite{Census2018}.

Ruktanonchai et al \cite{Ruktanonchai2016} estimate the local reproduction numbers $\R_{0,i}$ based on malaria incidence data reported in the Malaria Atlas Project, assuming that disease is at steady state.  We use their estimated local reproduction numbers for system \eqref{ModelEquations} here.   Values for the transmission parameters $\beta_i, \beta_{v,i}$ were then made assuming that these parameter values were equal within a given health district.

\begin{table}[ht!]
\centering{
\setlength{\tabcolsep}{2pt}
\begin{tabular}{ccc}\toprule
{Parameter} & {Value} & {Units} \\
\toprule
$\beta_{i}^{h}$ & $\big[0, 0.0279 \big]$ & days$^{-1}$\\ \vspace{.2cm}
$\mu_{i}^{h}$ & $\dfrac{1}{23002}$ & days$^{-1}$\\ \vspace{.2cm}
$\gamma_{i}^{h}$ &$\dfrac{1}{150}$ & days$^{-1}$ \\ \vspace{.2cm}
$\beta_{i}^{v}$ & $\big[0, 0.0279 \big]$& days$^{-1}$ \\ \vspace{.2cm}
$\mu_{i}^{v}$ &$\dfrac{1}{10}$ & days$^{-1}$   \\
\toprule
\end{tabular}
}
\caption{Parameters values for system \eqref{ModelEquations} used for the malaria Namibia case study.  Source: \cite{Ruktanonchai2016}.}
\label{ParameterValuesModel}
\end{table}

\subsection{Sensitivity analysis}

We now apply the sensitivity equations \eqref{EqSenBeta} - \eqref{EqSensFv} to the Namibia data.  Note that a necessary prerequisite for using the sensitivity equations is for the scale parameter $z$ to be within the radius of convergence of the Laurent series \eqref{eqn:laurent}, which is equivalent to  $1<\displaystyle\frac{1}{\rho(L^{d}D_{\mu+\gamma})}$.  This is indeed the case here, with $\displaystyle\frac{1}{\rho(L^{d}D_{\mu+\gamma})}=7.54$, well within the radius of convergence.

\begin{figure}[ht!]
    \begin{subfigure}[b]{0.5\textwidth}
        \includegraphics[width=\textwidth]{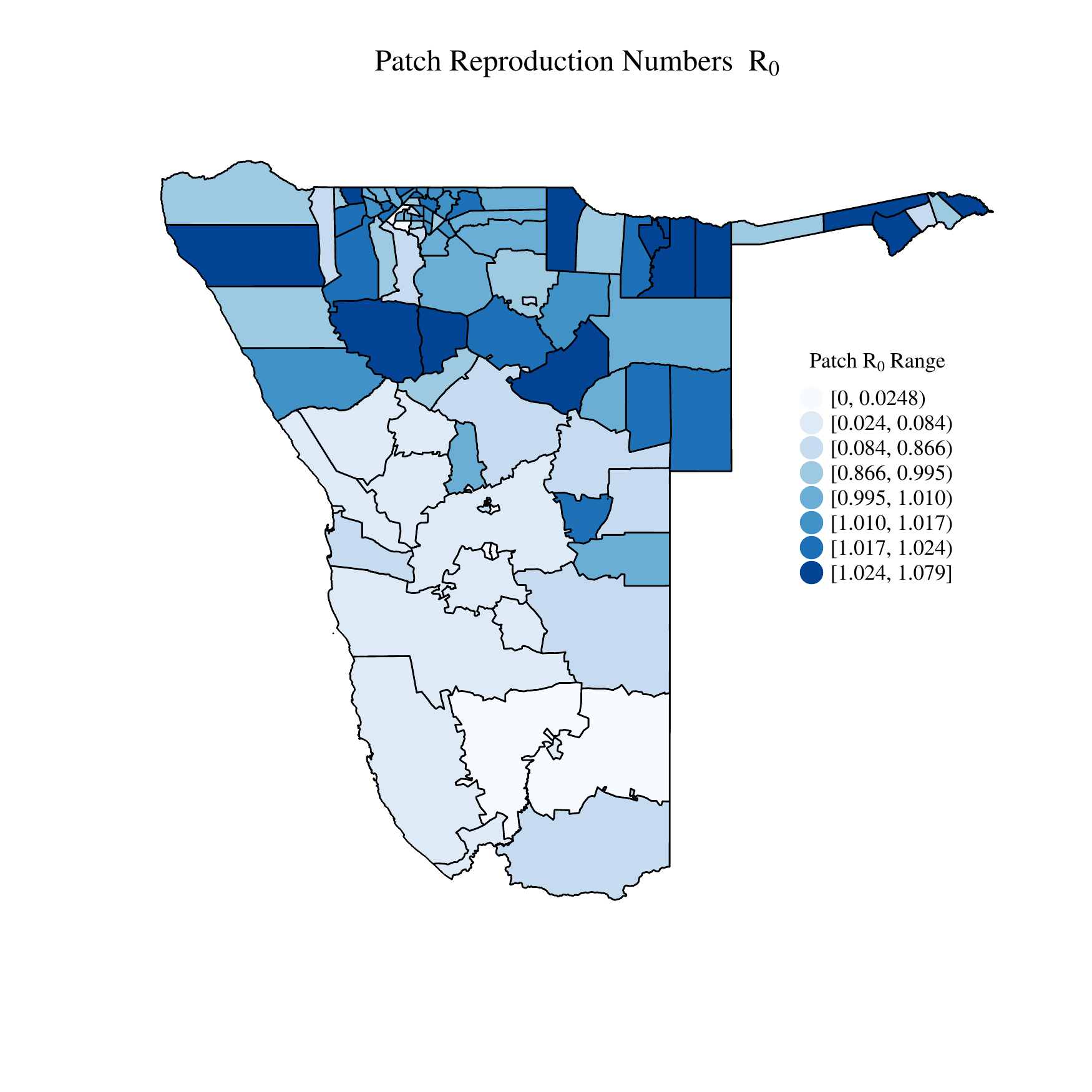}
        \caption{}
        \label{fig:R0map}
    \end{subfigure}%
    \begin{subfigure}[b]{0.5\textwidth}
        \includegraphics[width=\textwidth]{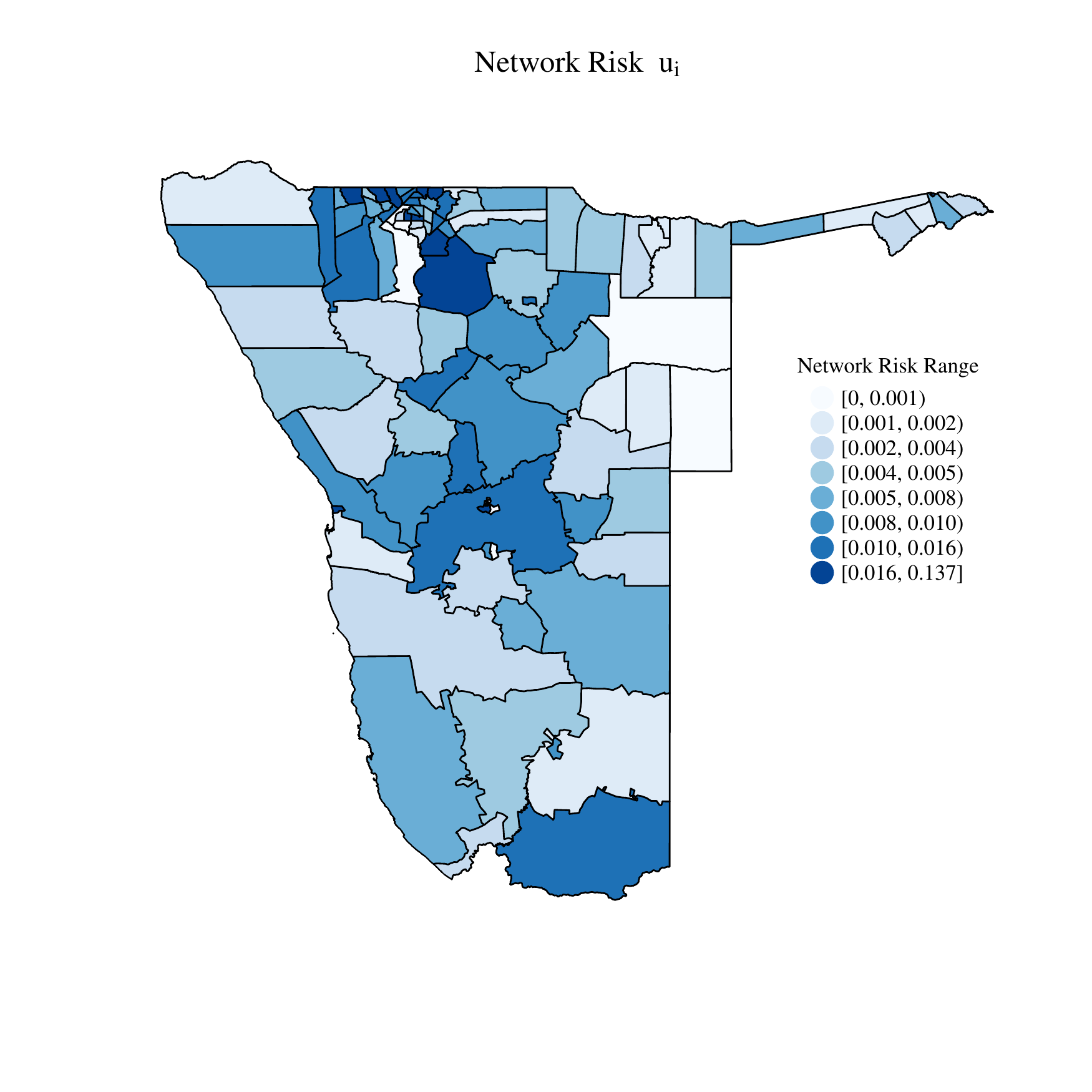}
        \caption{}
        \label{fig:umap}
    \end{subfigure}
    \caption{(a) Reproduction numbers for each health district in isolation, as estimated by \cite{Ruktanonchai2016}.  (b) Values for the network risk $u_i$ by health district. }
    \label{fig:R0_u_map}
\end{figure}

As noted in Section \ref{sect:sens}, the relative sensitivities between health districts depend upon $\R_{0,i} d_i u_i$ (for sensitivity to $\beta_i^{h}, \beta_{i}^{v},$ and $\mu_{i}^{v}$) and $\gamma_i^{h} u_i$ (for sensitivity to $\gamma_i$).  Here we use the same absorption rates $d_i$ and recovery rates $\gamma_i^{h}$ across health districts (Table \ref{ParameterValuesModel}), so the relative sensitivities are determined by the patch reproduction numbers $\R_{0,i}$ and the network structure through $u_i$.  Figures \ref{fig:R0map} and \ref{fig:umap} show $\R_{0,i}$ and $u_i$ by health district, respectively.

Comparison of patch reproduction numbers $\R_{0,i}$ and network risk $u_i$ in
Figure \ref{fig:R0_u_map} shows differences in which health districts correspond to transmission hot spots (high $\R_{0,i}$; Figure \ref{fig:R0map}) compared to health districts corresponding to mobility sinks (high $u_i$; Figure \ref{fig:umap}).  Transmission hot spots are found in rural health districts in the northeastern part of the country.  For example, the highest patch reproduction numbers are found in the Rundu Rural West,  Rundu Rural East, Mashare, and Mpungu health districts.  By contrast, mobility sinks are found in urban areas, with the highest values for $u_i$ corresponding to Windhoek West, Oshakati East, Rundu Urban, and Katima Muliro Urban.

\begin{figure}[h!]
    \begin{subfigure}[b]{0.5\textwidth}
        \includegraphics[width=\textwidth]{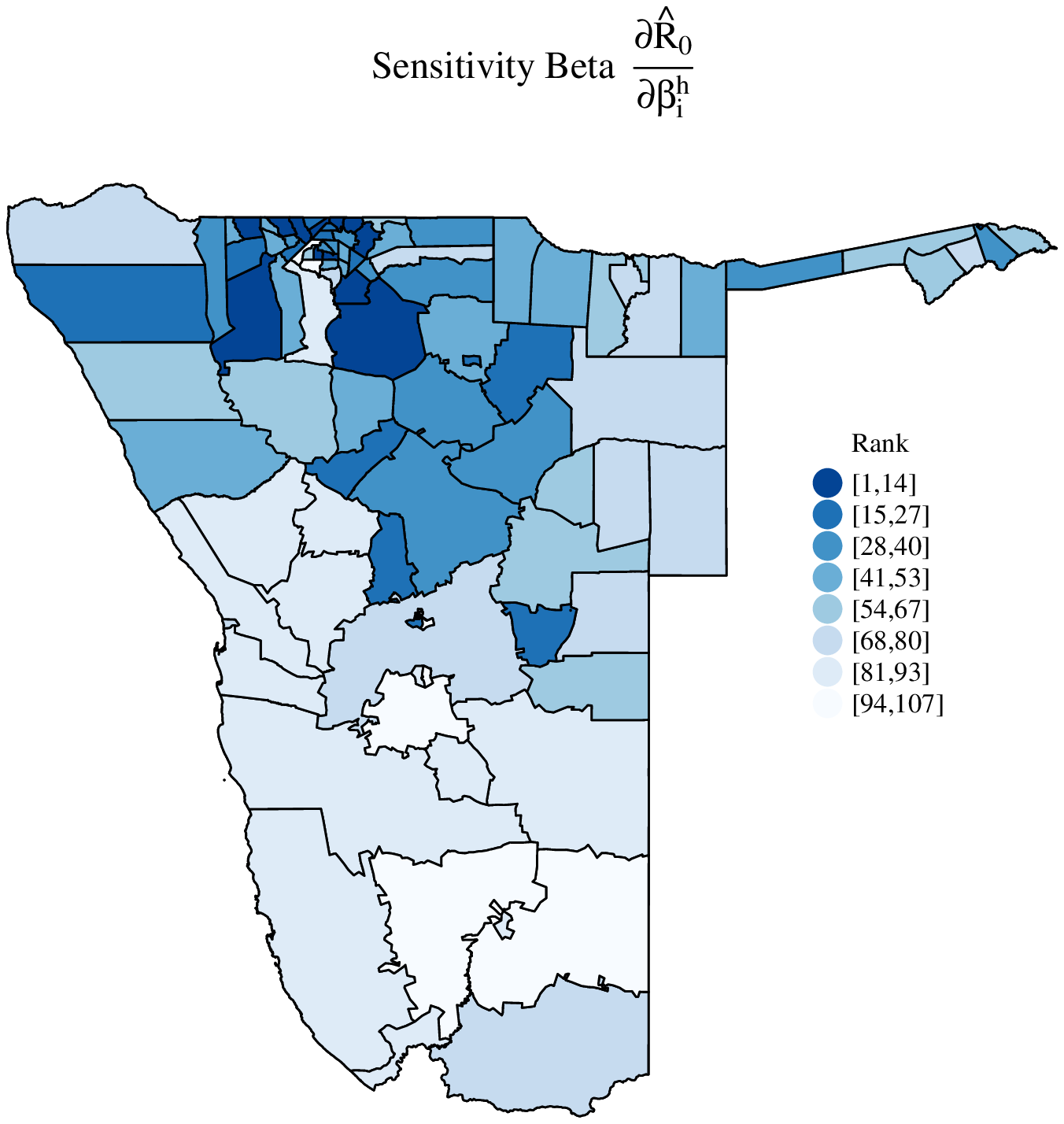}
        \caption{}
        \label{fig:map_beta}
    \end{subfigure}%
    \begin{subfigure}[b]{0.5\textwidth}
        \includegraphics[width=\textwidth]{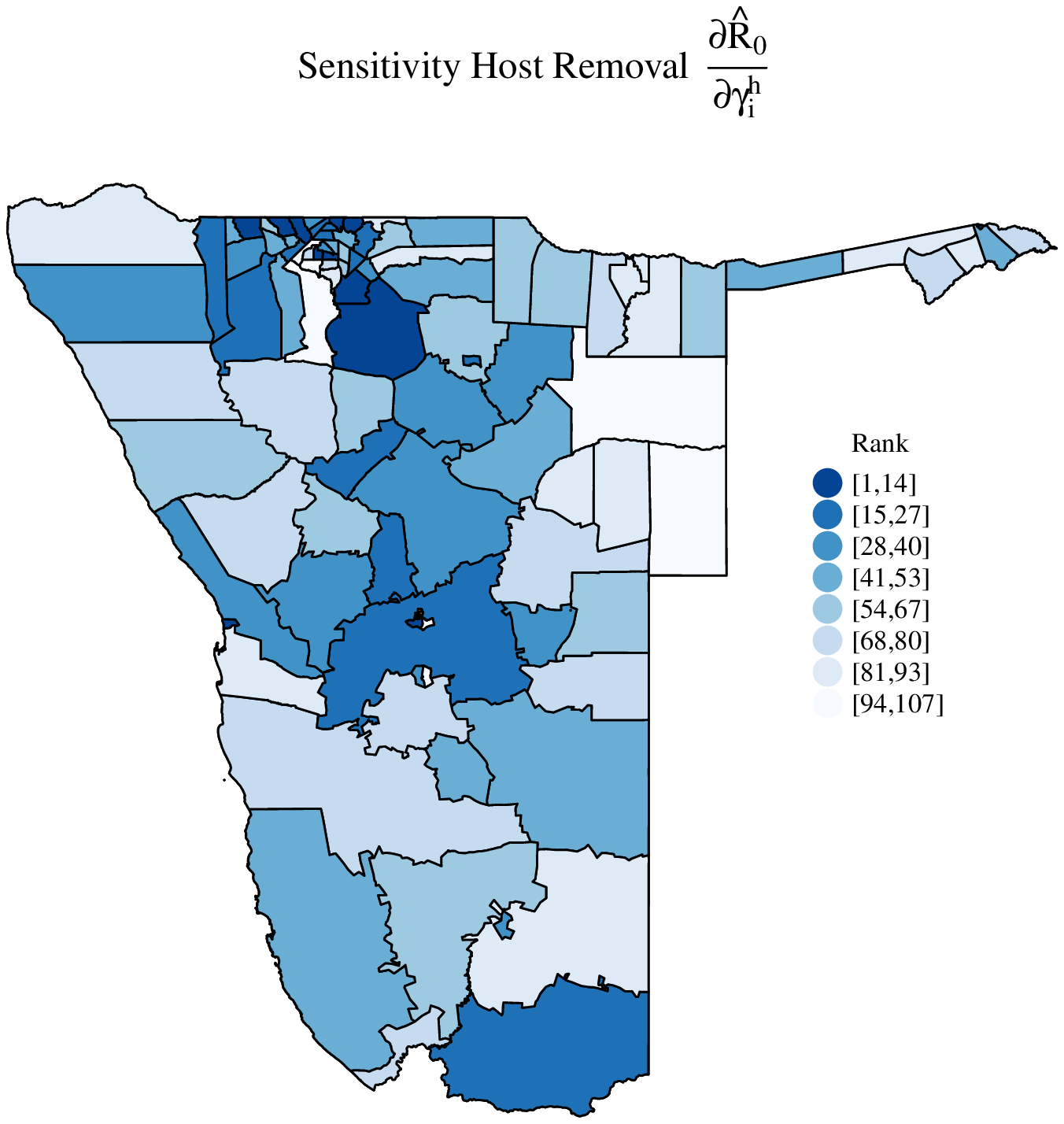}
        \caption{}
        \label{fig:map_gamma}
    \end{subfigure}
        \begin{subfigure}[b]{0.5\textwidth}
        \includegraphics[width=\textwidth]{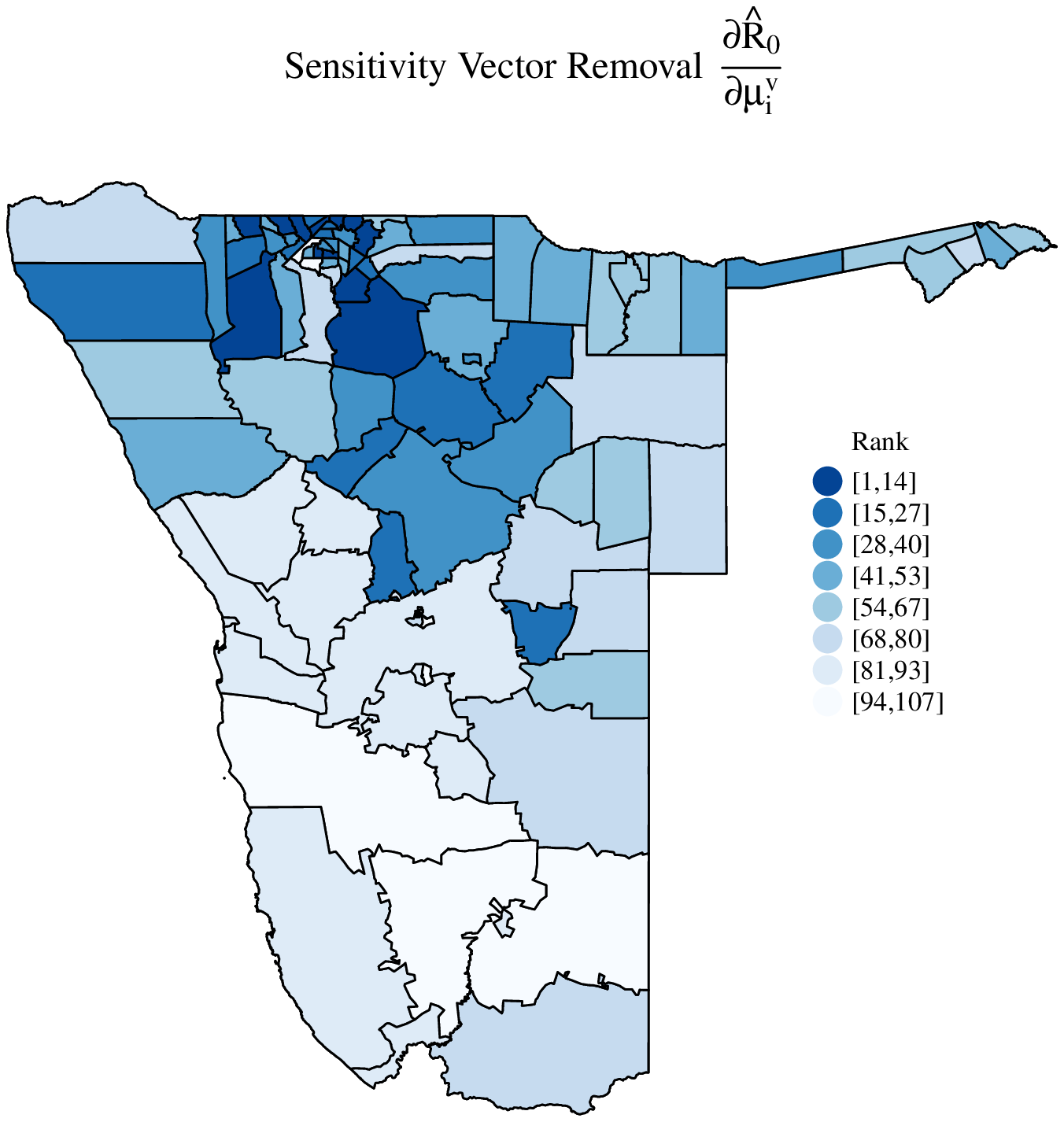}
        \caption{}
        \label{fig:map_fv}
    \end{subfigure}
    \caption{Comparison of $\hat{\R}_0$ sensitivities to health district parameters.  Color corresponds to rank (107=least sensitive health district, 1=most sensitive health district).
    (a) Sensitivities of $\hat{\R}_0$ to $\beta_i^{h}$, calculated from \eqref{EqSenBeta}.  (b) Sensitivities of $\hat{\R}_0$ to $\gamma_i^{h}$, calculated from \eqref{EqSensGamma}. (c) Sensitivities of $\hat{\R}_0$ to $\mu_{i}^{v}$, calculated from \eqref{EqSensFv}. }
    \label{fig:map_sens}
\end{figure}

Figure \ref{fig:map_sens} shows the relative sensitivities of $\hat{\R}_0$ to $\beta_i^{h}$ (Figure \ref{fig:map_beta}) and $\gamma_i^{h}$ (Figure \ref{fig:map_gamma}). As discussed in Section \ref{sect:toy}, the relative sensitivities of $\hat{\R}_0$ to $\beta_i^{h}, \beta_{i}^{v}$, and $\mu_{i}^{v}$ are determined by a combination of network structure and local transmission characteristics, via the $\R_{0,i} d_i u_i$ terms in \eqref{EqSenBeta} - \eqref{EqSensBetav} and \eqref{EqSensFv}.  For the Namibia data, the four health districts where the domain $\hat{\R}_0$ is most sensitive to these parameters are Oshakati East, Katima Muliro Urban, Rundu Urban, and Ondangwa.  Note that these health districts do not correspond to the four highest transmission hot spots, nor do they correspond to the ordering of the four highest mobility sinks (although Oshakati East, Katima Muliro Urban, and Rundu Urban are among the four health districts with highest $u_i$ values).  Thus intervention efficacy in terms of local transmission $(\beta_i^{h}, \beta_{i}^{v})$ and vector removal $(\mu_{i}^{v})$ is not determined by mobility sinks and transmission hot spots independently of one another, but upon a combination of these factors.  By contrast, the relative sensitivities to host recovery $\gamma_i^{h}$ depends only upon the network structure $u_i$ in this case study as we assumed the host recovery rates to be equal.

\begin{figure}[h!]
\centering
\scalebox{0.53}{\includegraphics{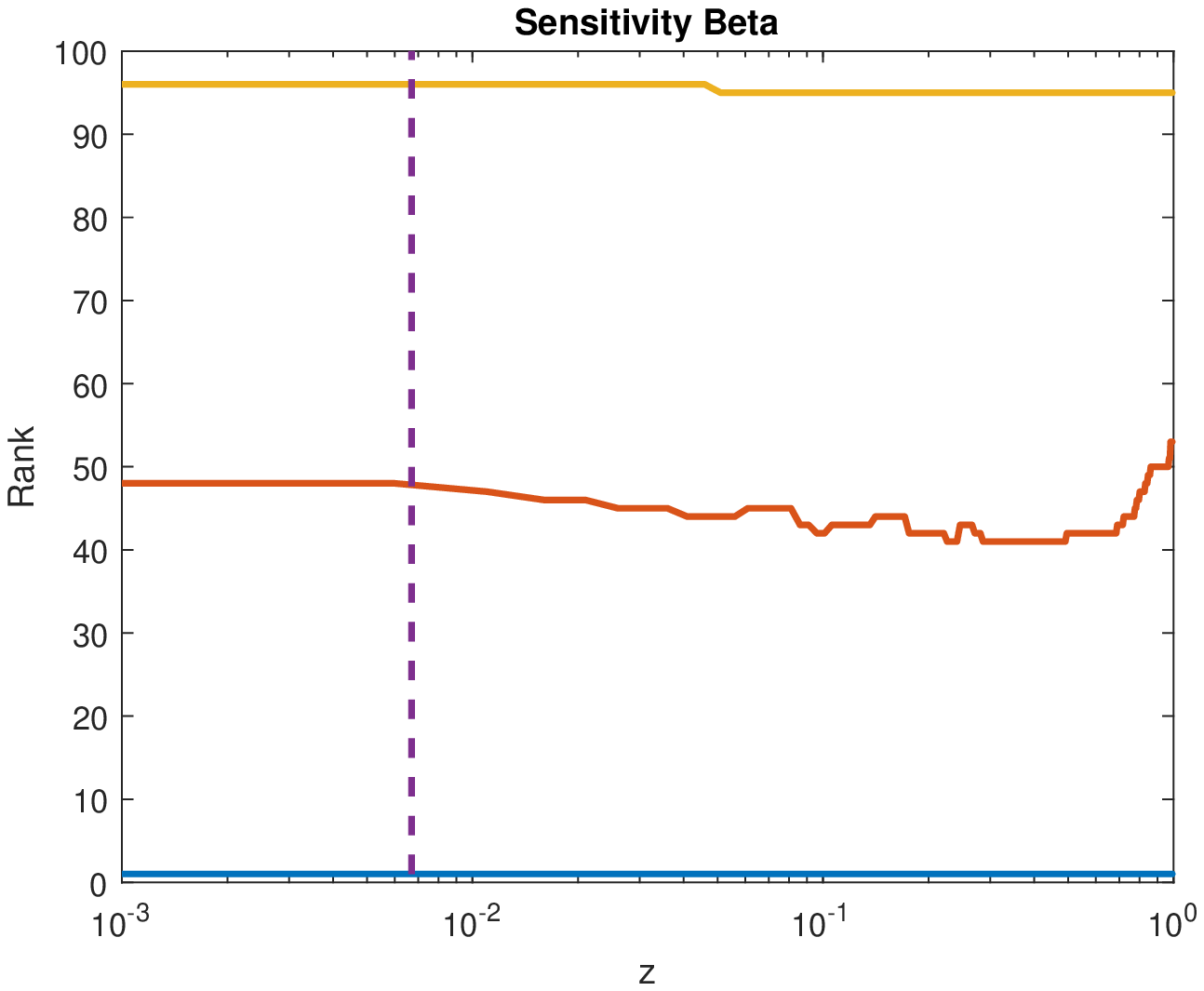}}
\scalebox{0.53}{\includegraphics{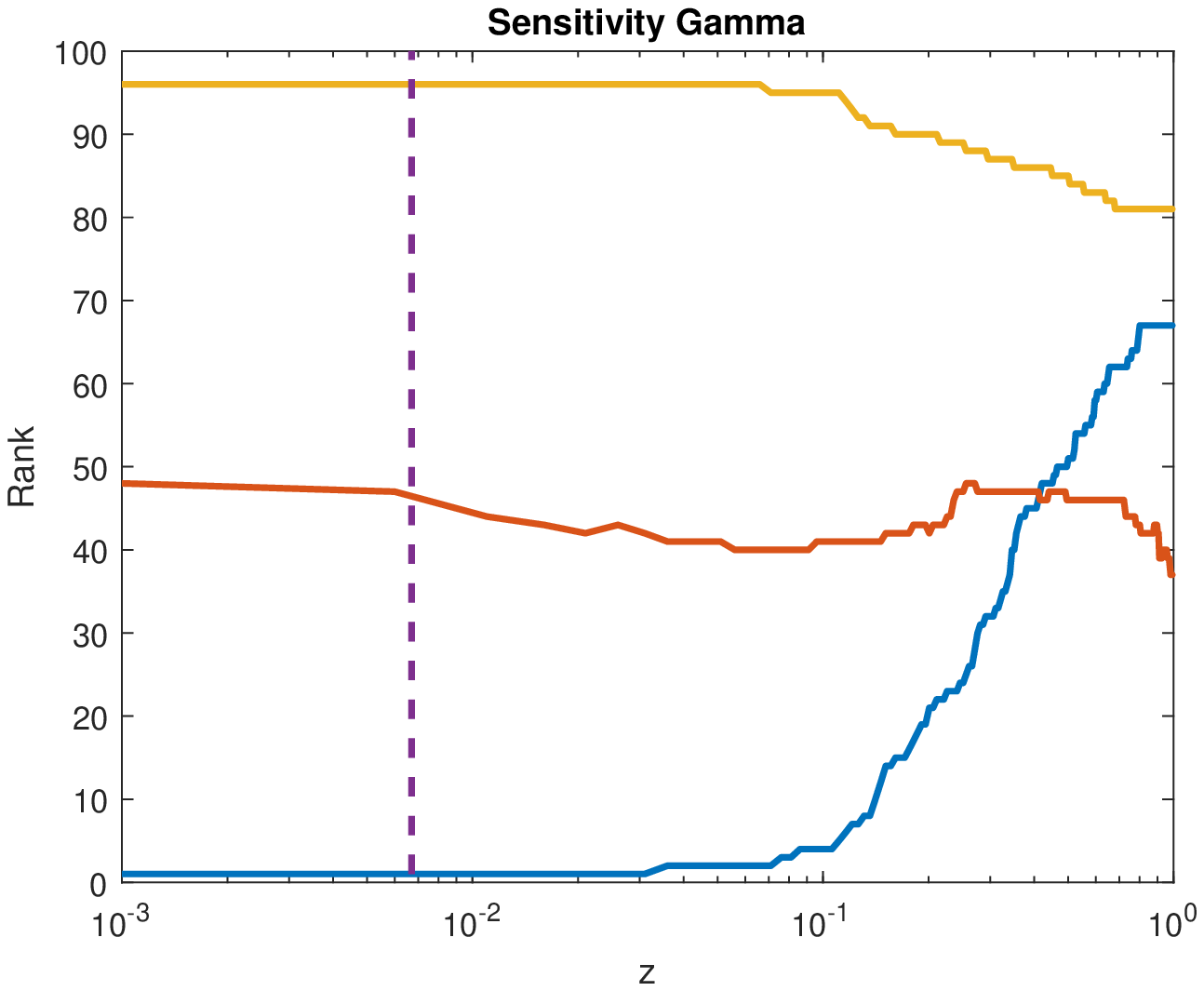}}
\scalebox{0.53}{\includegraphics{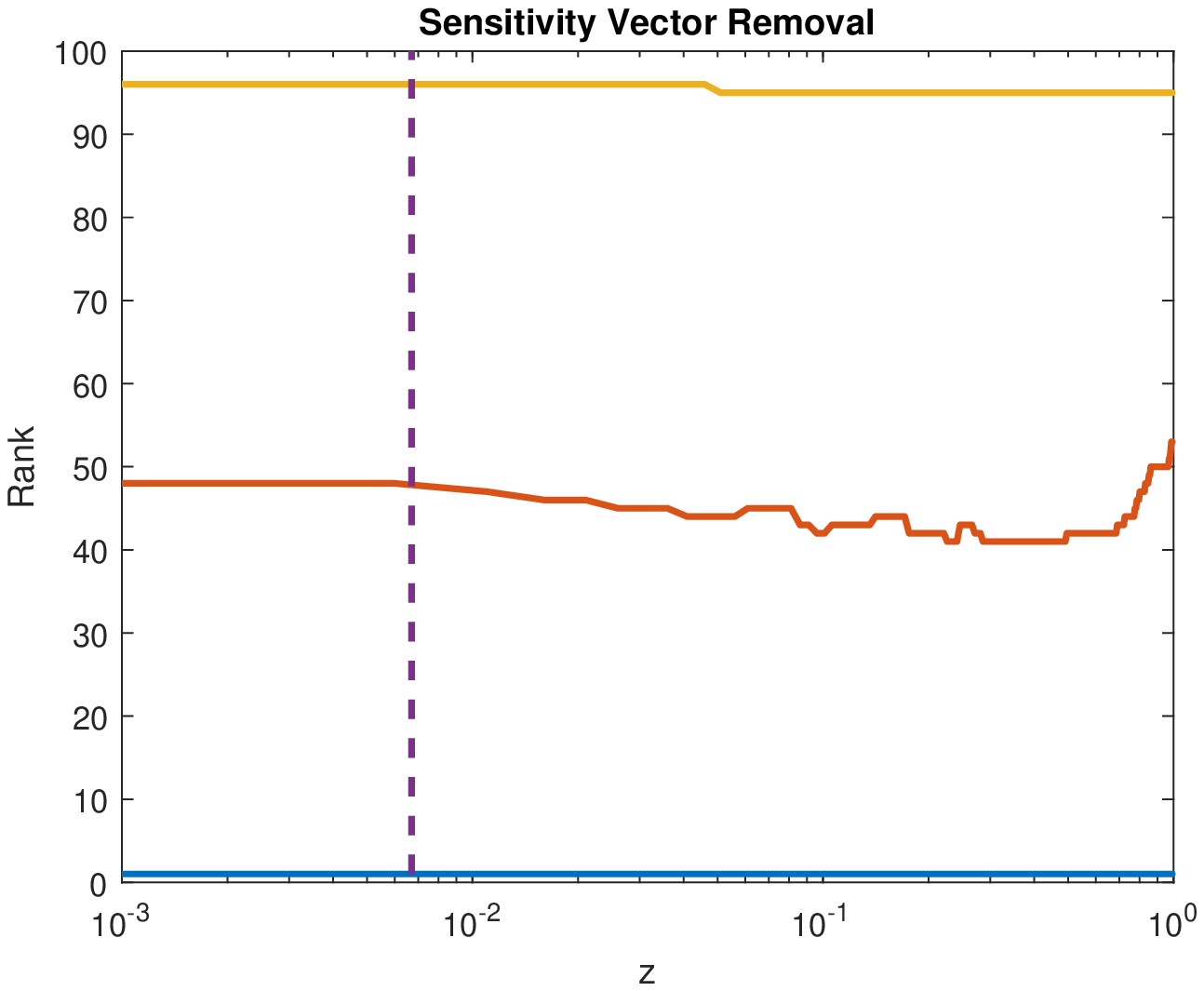}}
    \caption{ Rankings of the numerically computed sensitivities of $\hat{\R}_0$ for system (2.1) to patch parameters.  The yellow line represents the least sensitive node, the red represents the medium sensitive node, and the blue represents the most sensitive node. The purple dotted line represents the time scale from the data. As the time scale approaches zero, the simulated ranks converge to the analytic rankings.}
    \label{fig:Ranking graphs}
\end{figure}

The sensitivities shown in Figure \ref{fig:map_sens} were computed from \eqref{EqSenBeta}-\eqref{EqSensFv}, which in turn are based upon the lowest term in the Laurent series expansion \eqref{eqn:laurent}.  The accuracy of these sensitivities thus depends upon the scale parameter $z$ in \eqref{eqn:laurent}.  To examine the robustness of our conclusions from the sensitivities shown in Figure \ref{fig:map_sens}, we examine how numerically computed sensitivities vary over a range of values for $z$.  Figure \ref{fig:Ranking graphs} shows how numerically computed sensitivity rankings for $\beta_i^{h}, \gamma_i^{h},$ and $\mu_{i}^{v}$ change over a range of $z$ values. The rankings show very little change for $z<0.01$, which includes the estimated $z$ value of $0.0067$ from the Namibia cell phone data (dashed line).  These findings indicate that conclusions from the sensitivity equations \eqref{EqSenBeta}-\eqref{EqSensFv} are informative for the parameter ranges considered.

\section{Discussion}
\label{sect:discussion}

Disentangling the effects of network structure and transmission hot spots is a fundamental issue for understanding infectious disease dynamics.  The analytical results presented in this paper can help elucidate how network structure and local transmission characteristics combine to affect vector-host disease dynamics.  As our case study of malaria in Namibia illustrates, these analytical results can be applied to empirical disease settings of public health interest.  Closed form expressions approximating the domain $\R_0$ and sensitivities of $\R_0$ to model parameters such as those provided here are important beyond simply the numerical values they provide, as they give insight into the factors underlying $\R_0$ and the parameter sensitivities.

Data on both network connectivity as well as local transmission characteristics are needed to apply the methods presented in this work. The malaria case study in Namibia utilized cell phone data to estimate network connectivity, and disease prevalence data to estimate local transmission parameters.   Human mobility data are being collected from diverse sources, including traffic patterns, cell phone data, airflow, navigation app use, social media posts, and more.  Detailed data are being collected as well on host demography \cite{corder2020modelling} and vector habitat suitability \cite{ayala2009habitat,chahad2018effects}. Public repositories will facilitate research on integrating network and local transmission data for understanding disease dynamics.

The techniques presented here build off of methods developed in \cite{Tien2015} for a patch `SIWR' model \cite{tien2010} coupled by environmental pathogen movement.  The methodology is flexible and can be applied to a variety of epidemiological models.  This includes expansion of system \eqref{ModelEquations} to incorporate the movement of vectors, which is an important factor in considering vector-borne diseases \cite{Smith2004}.  The scales of movement for host and vector may differ significantly.   For example, the typical movement scale for mosquitoes is on the order of a few kilometers \cite{Kaufmann2004}.  Although there has been work that has incorporated host and vector movement \cite{Cosner2015, Phaijoo2017}, additional studies are needed that merge theoretical concepts with data to understand the interplay of the disease between the host and vector as seen here and in \cite{esteban2020}.

The Laurent series \eqref{eqn:laurent} underlying our analytical results converges for sufficiently small pathogen removal (`absorption') rates.  For the malaria in Namibia case study, this criterion is amply met.  Indeed, the estimated infectious period is sufficiently long that taking only the lowest order term in the series suffices to give a good approximation.  In other situations, additional terms may be needed. For example, a refinement of this work would be to directly estimate the local reproduction numbers for system \eqref{ModelEquations} from the Malaria Atlas Project incidence data. We note that for other disease settings, pathogen removal may be fast relative to movement.  In this `weak coupling' regime, a different series expansion for the fundamental matrix may be used.  Further study of how disease dynamics vary with coupling strength is an area for future work.

\section*{Acknowledgements}
The authors acknowledge the support of the Mathematical Biosciences Institute-DMS 1440386 and NSF grant-DMS 1814737. We would like to thank Dr. Nick Ruktanonchai for providing the mobility data that allowed us to conduct the case study in this article.


\bibliography{BibVector_rev4}
\bibliographystyle{elsarticle_num}

\end{document}